\documentclass[12pt,a4paper,jhep]{article}
 \pdfoutput=1
\usepackage{jheppub}
\usepackage{textcomp}

\usepackage{graphicx,subfigure,color,array,slashed,wasysym}
\usepackage{caption}
\usepackage{dcolumn}
\usepackage{bm}
\usepackage{hyperref}
\usepackage{amsmath,amssymb,amsbsy,amsfonts,latexsym,graphicx,hyperref}
\usepackage{bbold}

\newcommand{\be}{\begin{equation}}
\newcommand{\ee}{\end{equation}} 
\newcommand{\ba}{\begin{array}}
\newcommand{\ea}{\end{array}}
\newcommand{\bea}{\begin{eqnarray}}
\newcommand{\eea}{\end{eqnarray}}

\newcommand{\half}{\frac{1}{2}}

\newcommand{\vev}[1]{\langle #1 \rangle}

\usepackage{multirow}

\title{ \Large Ginzberg-Landau-Wilson theory for Flat band, Fermi-arc and surface states of strongly correlated systems} 
\author[a]{Eunseok Oh,}
\author[b]{Yunseok Seo,}
\author[a]{Taewon Yuk,}
\author[a]{Sang-Jin Sin}
\emailAdd{lspk.lpg@gmail.com}
\emailAdd{yseo@gist.ac.kr}
\emailAdd{tae1yuk@gmail.com}
\emailAdd{sjsin@hanyang.ac.kr}
\affiliation[a]{ Department of Physics, Hanyang University, Seoul 04763, Korea }
\affiliation[b]{ School of Physics and Chemistry, Gwangju Institute of Science and Technology, Gwangju 61005 Korea} 
 \abstract{  
 We consider a holographic theory  as a Ginzberg-Landau theory working for   strongly interacting  system near the quantum critical point:  we take  the   bulk matter field  $\Phi^I(r,x)$,  the dual of the  fermion bilinear, as  the  order parameter.   We calculate and classify the fermion spectral functions  in the presence of  such orders. Depending on the symmetry, we found spectral features  like the gap, pseudo-gap, flat disk bands and  the Fermi-arc connecting the two  Dirac cones, which are familiar in Dirac material and Kondo lattice.   
 Many of above features are associated with the zero modes  whose presence is tied with 
  a discrete symmetry of the  interaction. 
  The interaction induced zero modes either makes the   strongly correlated system fermi-liquid like,  or creates a  disk-like  flat band.   
Some of the order parameters  in  the bulk theory  do not have an interpretation of symmetry breaking  in terms of the boundary space, which opens the possibility of  'an order without symmetry breaking'. 
 }

\subheader{\today }
\arxivnumber{2007.xxxxx} 
\begin{document}
\maketitle
 
 \section{Introduction: holographic order parameter} 
The strong correlation is  property of a phase of general matters not a few special materials,  because even a weakly interacting material can become strongly interacting in some parameter region. It happens when the  fermi surface (FS) is tuned to be small, or when conduction band is designed to  be flat. The Coulomb  interaction in a metal is small  only  because the charge is screened by the particle-hole pairs which are abundantly created when FS is large.   In fact, any Dirac material  is strongly correlated as far as its FS is near the tip of the Dirac cone.  This was demonstrated in the clean graphene  \cite{pkim,Lucas:2015sya}  and  the surface of  topological insulator \cite{liu2012crossover,zhang2012interplay,bao2013quantum}  through the anomalous   transports   that could  be quantitatively explained by a holographic theory\cite{Seo:2016vks,Seo:2017oyh,Seo:2017yux}.     
In the cuprate and other transition metal oxides,  hopping of  the electrons in 3d shells are much slowed down because  the outermost  4s-electrons are taken by the Oxygen. In disordered system electrons  are slowed down by the Kondo physics\cite{Coleman:2015uma}. 
In twisted bi-layered graphene\cite{cao2018unconventional,cao2018correlated} flat band appears due to the formation of larger size effective lattice system called Moire lattice.  
In short,  strong correlation phenomena is ubiquitous,  where  the traditional methods are not working very well, therefore new method   has been longed-for for many decades.   

When the system is strongly interacting, it is hard to characterize the system in terms of its basic building blocks and one faces the question how to handle the huge degrees of freedom to make a physics, which would   allow just a few number of  parameters.
Recently, much interest has been given to the holography   as a possible tool for strongly interacting system (SIS) by applying the idea to describes the quantum critical point (QCP) describing for example the normal phase of unconventional superconductivity. 
Notice however that the QCP   is  often  surrounded by an ordered phase. Physical system can be identified  by the information of nearby phase as well as the QCP itself.  

 For the ordinary finite temperature critical point, the Ginzberg-Landau (GL)  theory  is introduced  precisely for that purpose. As is well known,  it  describes  the transition between the  ordered and disordered states near the   critical point. 
 It works for  weakly interacting theory and when it works   it is a simple but powerful.  The order parameter depends on the symmetry of the system and the phase transition is due to the symmetry breaking. 
 The tantalizing question is whether there is a {\it working} GL theory  for strongly interacting systems.   
The GL theory works also because of the universality  coming from the vast amount of information loss at   at the critical point, which resembles a black hole. 
For the quantum critical point, 
we need one more dimension to encode the evolution of   physical quantities along the probe energy scale  \cite{wilson1971renormalization,wilson1975renormalization}.      Therefore  it is natural to  interpret AdS/CFT \cite{Maldacena:1997re, Witten:1998qj,Gubser:1998bc}  as a GL  theory for the strongly interacting system where the radial coordinate   describe the dependence on the renormalization scale\cite{alvarez1998geometric,balasubramanian1999spacetime,de2000holographic,heemskerk2011holographic}.    
For this reason we call it as Ginzberg-Landau-Wilson theory. 

The transport and the spectral function (SF)  have been calculated in various  gravity backgrounds using the holographic method.  However, it has been  less clear  in general for what system such results   correspond to.   
For this we believe that the information on the ordered phase is as important as the information on the QCP itself. Clarifying this point will be the first step for more serious  condensed matter physics  application  of  the holography idea and this is the purpose of this paper.
 The idea is to introduce the holographic order parameters of various symmetry type and calculate the spectral function in the presence of the order. The resulting features of the fermion spectrum should  be compared with the Angle Resolved photo-emission spectroscopy (ARPES) data, which is the most important finger print of the materials. 
  
 Notice that both the magnetization and  the gap of superconductor can be understood as the expectation value of fermion 
bi-linears\cite{fradkin2013field} 
$\vev{\chi^\dagger \vec{\sigma} \chi}$ and $\vev{\chi   \chi}$ of the fermion $\chi$.
In fact,  the expectation value of any fermion bilinears  can play the role of leading order parameters.  
When two or more of them are non-zero, they can compete or coexist according to   details of    dynamics. 
Then, the most natural   order parameter in the holographic theory should be {\it the bulk dual field of the fermion bilinear} because it contains  the usual order parameter  as the coefficient of its sub-leading term  in the near boundary expansion. 
  The presence of the order parameter actually characterizes the physical system  off but near the critical point.  
We will calculate  spectral functions  \cite{sslee,Liu:2009dm,Iqbal:2009fd,Cubrovic:2009ye}   in the presence of the order parameter.
Our prescription for them is to add the Yukawa type interaction  between the  order parameter   and the fermion bilinear in the bulk and see its effect on the spectrum. 
\vskip .3cm
To be more specific, let $\psi_{0}$ be the source field of the fermion $\chi$ at the boundary and $\Phi_{0I}$ be the source of the fermion bilinear ${ \bar \chi}\Gamma^{I}\chi $ where $I=\{\mu_{1}\mu_{2}\cdots \mu_{n}\}$  represent different tensor types of Gamma matrix. 
The  extension of source fields $\psi_{0}$ and $\Phi_{0I}$  to the AdS bulk  is the bulk  dual field  $\psi$ and the order parameter field $\Phi_{I}$. 
We calculate the fermion spectral function by considering 
  the Yukawa type interaction of the form 
\be 
\Phi_I\cdot{\bar \psi}\Gamma^I\psi. 
\ee  
For example, the complex scalar can be associated with the  superconductivity, and the neutral scalar  to a magnetic order.  
We will  classify 16  types  of interactions into a few class of scalars, vectors and  two-tensors  and calculate the spectral functions.   With such  tabulated results, one may identify the order parameter of a physical system  by comparing the ARPES data with the spectral functions.   

Some of the idea has been explored for   scalar  \cite{Faulkner:2009am} and  tensors \cite{benini2011holographic,Vegh:2010fc}  to discuss  the spectral gap of the   superconductivity.  But in our paper, we will see much more variety of spectral features like flat band, pseudo gap, surface states, split cones and nodal line etc. 
 The most studied  feature of the fermion spectral function is the gap. The authors of    \cite{Edalati:2010ww,Edalati:2010ge} considered   the dipole term ${\bar\psi}F_{rt}\Gamma^{rt}\psi$ to discuss the  Mott  gap.  However, if we define the gap as vanishing density of state for a finite width of energy around the fermi level, the dipole term does not generate  such spectrum because the band created by the dipole interaction approaches to the Fermi level for large momentum. In \cite{Vegh:2010fc} the author  reported the observation of   Fermi-arc in the sense of incomplete Fermi surface.  Our Fermi-arc is in the sense of surface state in the presence of  various different  types of vector order. 
 
One of the surprising aspects of our result is that usual scalar interaction creates not a gap but a vivid zero modes which were absent without the order parameter coupling. 
 We find that the pseudo scalar interaction    $ \Phi_{5} {\bar\psi}\Gamma^{5}\psi$   generates a gap as it was discovered in \cite{Faulkner:2009am}.   We found that  the parity symmetry  controls the presence of the zero mode.  
 Another interesting aspect is that 
 some of the order parameters  in holographic theory, especially those of tensors with radial index  do not have direct symmetry breaking interpretation in the boundary theory, and this   opens the possibility of  'an order without symmetry breaking'. 




\section{Flat spacetime spectrum for various Yukawa interactions} 
To learn the effect  of  the each type of interaction, 
we first study the spectral functions(SF) of  flat space fermions and classify them.  
The spectral functions will be delta function sharp. This will help us by suggesting  what  to expect in curved space if there are correspondence, because the AdS version will be a deformed and blurred version of flat space SF by interaction effects which is transformed into the geometric effect.  
However, AdS$_{4}$ and its boundary has difference in the number of independent gamma matrices,   threrefore there are interaction terms in the bulk which does not have analogue in its boundary fermion theory. 

We now consider boundary fermion $\chi_{1}$, $\chi_{2}$ whose action is given by 
\bea
&S=S_{ \chi}+S_{\Phi}+S_{int},  \quad {\rm where} \\
		& S_{ \chi} =\int d^{3}x\; \sum_{j=1}^{2}i\bar{ \chi}_j
		\gamma^\mu\mathcal{D}_\mu \chi_j  \\
		& S_{\Phi}=\int d^{3}x\left((D_{\mu}\Phi_{I})^2 -m^{2}_{\Phi} \Phi_{I}\Phi^{I}) ,\right) \\
		& S_{int}=p_1\int d^{3}x\left(\bar{ \chi_1}\,\Phi\cdot\gamma\, \chi_1+h.c \right) + p_{2}\sum_{j=1}^{2}\int d^{3}x\left( \bar{ \chi_1}\, \Phi\cdot\gamma\, \chi_2+h.c\right) ,
			\eea
where $\Phi\cdot\gamma=\gamma^{\mu_{1}\mu_{2}\cdots\mu_{I}}\Phi_{\mu_{1}\mu_{2}\cdots\mu_{I}}$ and $I$ is the number of the indices. 
For one flavor case,  we set $p_{1}=1, p_{2}=0$ and 
	   set $p_1=0, p_{2}=1$ for 2 flavor. 
	  Each two component fermion in 2+1 dimension has definite helicity  and the spin is locked with the momentum. 	Therefore with one flavor, we can not have a Pauli paramagnetism. 
	We list $2\times 2$  gamma matrices of 2+1 dimension. 
	\begin{align}
		& \gamma^t=i\sigma_{2},\quad \gamma^x=\sigma_{1}, \quad \gamma^y=\sigma_{3}, 
		 \\
		& \gamma^{\mu\nu}=\frac{1}{2}\left[\gamma^\mu,\gamma^\nu\right], \quad \gamma^{tx}=\sigma_{3}, \quad \gamma^{ty}=-\sigma_{1}, \quad \gamma^{xy}=-i\sigma_{2} 
	\end{align}
Following identity is necessary  and useful to construct  lagrangian. 
\be
\gamma^{\mu\dagger}=\gamma^{0}\gamma^{\mu}\gamma^{0} ,\quad 
\hbox{ and }\quad 
\gamma^{\mu\nu}=\epsilon_{\mu\nu\lambda}\gamma_{\lambda}, 
\label{2p1}
\ee	  

\subsection{Spectrum in flat space}
Because we did not introduce  a lattice structure, we do not have periodic structure in momentum space. instead
we focus on the band structure near the zero momentum. If we include only one flavor,   only two bands will appear in the spectrum.  For the zero mass, left and right modes can be split, while it can not be for the massive case. 
For two flavors, the number of bands is just doubled. 

\subsubsection{One flavor case: $\bar{ \chi_1}\,\Phi\cdot\gamma\, \chi_1$}
\subsubsection*{Scalar : $\Phi\cdot\gamma=\Phi$}
For the flat space, there is not much  difference between the   scalar interaction and the mass term. Gap is generated as one can see from the equation of motion. See also the figure \ref{flatspec}(a).  The mass term, if exist,  violate the parity symmetry.  	 
\subsubsection*{Vector : $\Phi\cdot\gamma=\boldmath{B_{\mu}}							\gamma^{\mu}$}
\begin{figure}[ht!]
			\centering
			\subfigure[Int. with $\Phi$]
			{\includegraphics[width=3cm]{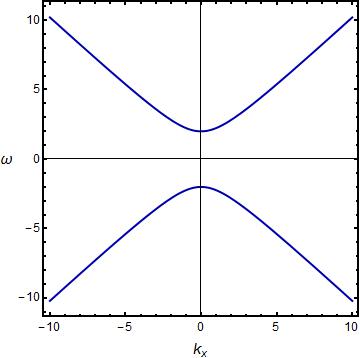}}
		\subfigure[ Int.with $B_{t}$]
			{\includegraphics[width=3cm]{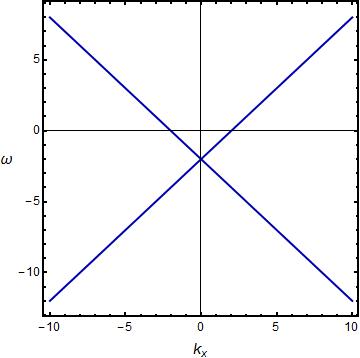}}
				\subfigure[ Int.with $B_{i}$]
			{\includegraphics[width=3cm]{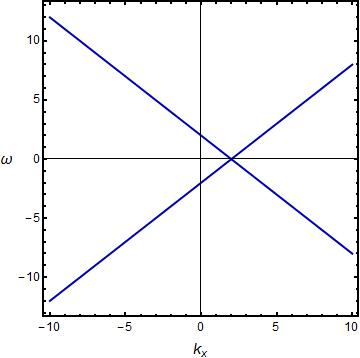}}
\caption{SF for one flavor. (a) scalar Interaction generate a gap.	  (b) $B_t$   shift the spectrum along $\omega$ direction.   $B_t=2$			 (c) $B_i$   shift the  spectral cone in $k_{i}$ direction. $B_i=2$
			}\label{flatspec}
		\end{figure}

	Its  effect   is  shifting the spectral cone in $x^{\mu}$ direction. See  figure \ref{flatspec}(b,c). 
 	\subsubsection*{Antisymmetric Tensor : $\Phi\cdot\gamma=\boldmath{B_{\mu\nu}}\gamma^{\mu\nu}$}
In 2+1, 
The role ofInt. $B_{\mu\nu}$ is  the same as  that of $\epsilon_{\mu\nu\lambda}B^{\lambda} $ due to the second identity of eq.(\ref{2p1}). Therefore  no new spectrum is  generated. 
 		\begin{figure}[ht!]
			\centering
			\subfigure[Int.with $\Phi$]
			{\includegraphics[width=3cm]{./fig/figfs/M/1flavor/kx/M_2_.jpeg}}
			\subfigure[Int.with $B_{t}$]
			{\includegraphics[width=3cm]{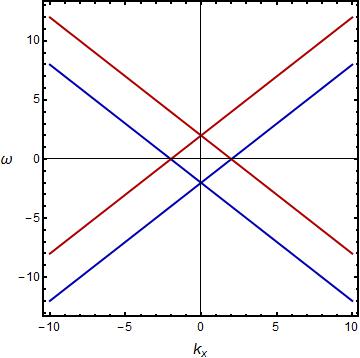}}
			\subfigure[Int.with  $B_{i}$]
			{\includegraphics[width=3cm]{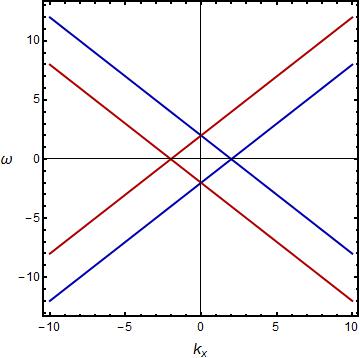}}
			\caption{SF for two flavors. (a) scalar interaction  generates a gap. 
			(b) $B_t$   shift the spectrum along $\omega$ direction. The configuration has rotational symmetry in $k_{x},k_{y}$ space.   
 			 (c) $B_i$   shift the  spectral cone along $k_{i}$ direction. 
			 Different flavor shifts in opposite direction. 
		}
		\end{figure}
Comparing with Figure 1 and Figure 2, we can see that the spectral double of the two  flavor case is manifest as doubling of the bands.

\subsubsection{2 flavor: $\bar{ \chi_1}\,\Phi\cdot\gamma\, \chi_2+h.c$}
Here  for convenience, we consider   parity  symmetry invariant   combination of interaction terms, 	
\begin{itemize}
\item  Scalar : ${\cal L}_{int}= i\Phi({\bar\chi}_{1}\chi_{2}+{\bar\chi}_{2}\chi_{1})$ or  $ \Phi({\bar\chi}_{1}\chi_{2}-{\bar\chi}_{2}\chi_{1})$ .
\item{Vector :  ${\cal L}_{int}= \boldmath{B_{\mu}} ({\bar\chi}_{1}\gamma^{\mu}\chi_{2}+{\bar\chi}_{2}\gamma^{\mu}\chi_{1})$, or   $\boldmath{iB_{\mu}} ({\bar\chi}_{1}\gamma^{\mu}\chi_{2}-{\bar\chi}_{2}\gamma^{\mu}\chi_{1})$ }
\item{Antisymmetric Tensor : ${\cal L}_{int}= \boldmath{B_{\mu\nu}} ({\bar\chi}_{1}\gamma^{\mu\nu}\chi_{2}+{\bar\chi}_{2}\gamma^{\mu\nu}\chi_{1})$, or $\boldmath{iB_{\mu\nu}} ({\bar\chi}_{1}\gamma^{\mu\nu}\chi_{2}-{\bar\chi}_{2}\gamma^{\mu\nu}\chi_{1})$
}
\end{itemize}
The point is that when the order parameter fields has non-zero vacuum expectation values,  the result of the operation 
 depends on the 
fluctuating fields, the interactions are invariant but when 
In each case, two forms of the interaction are equivalent because 
the second form is just unitary transform of the first   by $\chi_{1}\to -i\chi_{1}, \chi_{2}\to \chi_{2}$.   Notice that  when the equation motion does not involve the $i$ in the interaction term,  the flavors shift in opposite direction for vector and anti-symmetric tensor cases while two flavors share the same spectrum for the scalar interaction. 

The spectrum for the two flavor system  is   a  double of  one flavor case.  
For scalar, gap is generate  and spectrum is degenerated because two flavors has identically gapped spectrum.   
For vector interaction,   the spectral cone of each flavor is shifted  in opposite direction.  
 	Therefore in 2+1 dimensional flat space,  anti-symmetric sector 
	can be mapped to the vectors. 
	because the role of $B_{\mu\nu}$ is that that of $\epsilon_{\mu\nu\lambda} B^{\lambda}$. 
However, in anti-de Sitter space, two sectors  can be different. 

\section{The fermions   in $AdS_{4}$ }
\subsection{   Dirac fermions   in flat 2+1 space  and in  $AdS_{4}$ }	
For massless case, the spin-orbit coupling locks the spin direction to that of the momentum so that for fixed momentum
 only one helicity is allowed for one flavor. In AdS$_{4}$, half of the  fermion components are projected out depending on  the choice of the boundary terms\cite{laia2011holographic}.   The spectrum  of the fermions with AdS bulk mass term is still   gapless, unless interaction creates a gap, because the  AdS bulk mass is a measure of the scaling dimension not a gap.   Therefore  4-component AdS$_{4}$ fermion suffer the same problem of 2 component massless fermions in 2+1 dimension. For example such spin-momentum locked fermion system does not have a Pauli paramagnetism\cite{Alexandrov:2012xe}. One way to avoid such problem is to introduce two flavor and create a gap in the   spectrum by coupling with non-zero  scalar  field $\Phi$ as we will show later. 
 
 A Dirac fermion in real system is that of  3+1 dimension even in the case the system is arranged into a two dimensional array of atoms. Therefore it should be described by two flavor of two component  fermions, which  corresponds  to two flavor 4-component fermions in AdS. Then the spectrum of massless Dirac fermion   in condensed matter system should be described as a degenerated   Dirac cones. To describe the sublattice  structure of the graphene, we need another doubling of the flavor.  
Therefore we consider only two flavor cases in the maintext,   and provide  the spectrum of the one flavor in the appendix for curiosity.  

 Notice that in 2+1 dimension, a fermion field has two component while in AdS$_{4}$ it has 4 components, where     only half of the fermion components are physical\cite{laia2011holographic}. Therefore,    the degrees of freedom of the bulk match with those of boundary     in $AdS_{4}$ theory if the number of flavor in each side are the same. However,  in $AdS_{5}$,   we need to  double the number of the fields, because 4 components in the boundary corresponds to the 8 components in the AdS bulk. 
  To avoid too many cases,  we will consider only $AdS_{4}$ cases  here, and treat the AdS5 separately in the future if necessary. 
The boundary action must be chosen such that it respect the Parity symmetry as we have done in eq. (\ref{bdryaction}), otherwise the flat space and curved space does not have correspondence especially in scalar order. 
  
\subsection{ Fermion action and equation of motion}
We consider the action of  bulk fermion $\psi$ which is the dual to the boundary fermion $\chi$.  
Let $\Phi^{I}$ be the  dual bulk field of the operator ${\bar \chi}\Gamma^I\chi$. 
The question is how the $\Phi^{I}$ couples to the bulk fermion $\psi$.  
When $\Phi^{I}$  is a complex field, it describe a charged order  like the  superconductivity that has been already studied in holographic context.\cite{Hartnoll:2008vx,Gubser:2008px}. 
If   it is real, it describes   a magnetic order  like anti-ferromagnetism or gapped singlet order. 
The main difference is the absence or presence of the 
order parameter with vector field $A_{\mu}$ which is dual to the electric current $J^{\mu}$. We will consider both cases simultaneously and summarize simply as ``without or with chemical potential'',  $\mu=A_{t}(r)|_{=\infty}$. 

 The action is given by the sum $S=S_{ \psi}+S_{bdry}+S_{\Phi}+S_{int},$ where 	
 \begin{align} 
		& S_{ \psi} =\int d^{4}x\; \sum_{j=1}^{2}i\bar{ \psi}_j
		\gamma^\mu\mathcal{D}_\mu \psi_j  
-im({\bar\psi}_{1}\psi_{1}-{\bar\psi}_{2}\psi_{2}), \label{action}\\
		& S_{ bdry}=\half \int_{bdry} d^{3}x\; i({\bar\psi}_{1}\psi_{1}+{\bar\psi}_{2}\psi_{2})\label{bdryaction},\\
			& S_{\Phi}=\int d^{4}x\sqrt{-g}\left(|D_{\mu}\Phi_{I}|^2 -m^{2}_{\Phi} \Phi^{*}_{I}\Phi^{I}\right), \\
		& S_{int}=p_{2f}\sum_{j=1}^{2}\int d^{4}x\left( \bar{ \psi_1}\, \Phi\cdot\gamma\, \psi_2+h.c\right)+ p_{1f}\int d^{4}x\left(\bar{ \psi_1}\,\Phi\cdot\gamma\, \psi_1\right)  ,
			\end{align}
where $\Phi\cdot\gamma=\gamma^{\mu_{1}\mu_{2}\cdots\mu_{I}}\Phi_{\mu_{1}\mu_{2}\cdots\mu_{I}}$ and it is important to remember that for scalar $\gamma\cdot\Phi=i\Phi$.
For one flavor,  $p_{2f}=0$ and 
	for 2 flavor,   $p_{1f}=0$. 
	 Also depending on real/complexity of $\Phi^{I}$,  the covariant derivative $D_{\mu}=\partial_{\mu}-igA_{\mu}$ has $g=0$ or 1, and 
we   use the AdS Schwarzschild or Reisner-Nordstrom      metric.    
\begin{align}
	d s^2&=-\frac{r^2}{L^2}f(r)d t^2+\frac{L^2}{r^2f(r)}d r^2+\frac{r^2}{L^2}d x^2\nonumber
	\\
	f(r)&=1-\frac{{r_H}^3}{r^3}-\frac{r_H \mu^2}{r^3}+\frac{r_H^2 \mu^2}{r^4}
\end{align}
where the horizon of the metric $r_H=\frac{1}{3}(2 \pi T + \sqrt{4\pi^2 T^2+3 \mu^2})$ and $\mu$ is a chemical potential.  

 Following the standard dictionary of AdS/CFT for the $p$-form bulk field $\Phi$ dual to the operator $O$ with dimension $\Delta$, its mass is related to the operator  dimension by 
\be 
m^{2}_{\Phi}=-(\Delta-p)(d-\Delta-p),
\ee
and asymptotic form near the boundary is 
\be 
\Phi=\Phi_{0}z^{d-\Delta-p}+ \vev{O_{\Delta}}z^{\Delta-p} .
\ee 
For the $AdS_{4}$,  $d=3, p=2, \Delta=2[\psi]=2$, we    should set  
 	  \be
	m_\Phi^2=0, \quad {\rm and }\quad   B_{\mu\nu}=B_{\mu\nu}^{(-1)} z^{-1} +B^{(0)}_{\mu\nu} .
	  \ee
 Here we used the coordinate $z=1/r$ which is simpler due to  the homogeneity of the AdS metric in this coordinate.  We can find out the expression of the fields in $r$ coordinate by using the tensorial property. 
%

Throughout this paper, we use  the    probe  solution $\Phi$ which is the solution in the pure AdS background. This approximation can give qualitatively the same behavior of the fermion spectral function because for  finite temperature, the horizon of the black hole cut out the black hole's inner  region where the true solution of $\Phi$ deviate much from the probe solution.   
  
Following \cite{Liu:2009dm}, we   introduce  $\phi_{\pm}$ by 
\bea
\psi_\pm = {(-gg^{rr})}^{-\frac{1}{4}} 
 \phi_\pm, \quad \phi_\pm = 
\left( 
  y_\pm    ,
  z_\pm     
\right)^T. \label{Eq:psi_pm}
\eea 
Then the  equations of motion  for the  one flavor, with all the possible terms turned on,  can be  written as 
\begin{align}
(\partial_r+\mathbf{U_K})\phi +\mathbf{U_I}\phi=0, \quad
\phi=\left( 
y_{+},     
z_{+},   
y_{-},   
z_{-}    \right)^T
\end{align} 
where matrix $U_K$ is from the kinetic terms and $U_Y$ is from the   interaction term. If all types of interaction terms are turned on,  they are given by 

\begin{align}
\hskip -2cm \mathbf{U_K}=& -i \frac{\omega}{r^2f}\Gamma^{rt}+i \frac{k_{x}}{r^2\sqrt{f}}\Gamma^{rx}+i \frac{k_{y}}{r^2\sqrt{f}}\Gamma^{ry}
-i g \frac{A_{t}}{r^{2}f}\Gamma^{rt}
- \frac{m}{r\sqrt{f}}\Gamma^{r}  , \qquad {\rm and} \\
\mathbf{U_I}=&- \frac{\Phi}{r\sqrt{f}}\Gamma^{r}-i \frac{\Phi_5}{r\sqrt{f}}\Gamma^{r5} +\frac{B_{xy}}{r^{3}\sqrt{f}}\Gamma^{t5} +i \frac{B_{rt}}{r\sqrt{f}}\Gamma^{t}+i \frac{B_{rx}}{r}\Gamma^{x}\nonumber\\ \nonumber
&+i \frac{B_{ry}}{r}\Gamma^{y}-\frac{B_{tx}}{r^{3}f}\Gamma^{y5} +\frac{B_{ty}}{r^{3}f}\Gamma^{x5} -i \frac{B_{x}}{r^{2}\sqrt{f}}\Gamma^{rx} \nonumber
\\&-i \frac{B_{y}}{r^{2}\sqrt{f}}\Gamma^{ry}-i \frac{B_{t}}{r^{2}f}\Gamma^{rt}-i B_{r}\mathbb{1}- \frac{B_{5x}}{r^{2}\sqrt{f}}\Gamma^{ty}+\frac{B_{5y}}{r^{2}\sqrt{f}}\Gamma^{tx} \nonumber
\\&-\frac{B_{5t}}{r^{2}f}\Gamma^{xy}-i B_{5r}\Gamma^{5}, 
\end{align}
where 
\bea
\Phi &=\frac{\Phi_{(s)}}{r}+\frac{\Phi_{(c)}}{r^2}, \quad  \Phi_5&=\frac{\Phi_{5(s)}}{r}+\frac{\Phi_{5(c)}}{r^2} \nonumber\\
B_{\mu\nu}&=r B_{\mu\nu(s)}+B_{\mu\nu(c)},\quad 
B_{r\mu}&=\frac{B_{r\mu(s)}}{r}+\frac{B_{r\mu(c)}}{r^2} \nonumber
\\
B_{\mu}&=B_{\mu(s)}+\frac{B_{\mu(c)}}{r}, \quad 
 B_{5\mu}&=B_{5\mu(s)}+\frac{B_{5\mu(c)}}{r},
 \nonumber \\
 B_{r}&=\frac{B_{r(s)}}{r^2}+\frac{B_{r(c)}}{r^3},\quad 
B_{(5)r}&=\frac{B_{5r(s)}}{r^2}+\frac{B_{5r(c)}}{r^3}, \nonumber
\eea 
where the index $i,j$ runs $t,x,y$ and $f$ is the screening factor of the metric. 
For AdS Schwartzschild case, $f=1-r_{H}^{3}/r^{3}$. 

For two flavors,  the equation of motion changes minimally: 
  \begin{align}
(\partial_r+\mathbf{U_K})\phi_{1} +\mathbf{U_I}\phi_{2}&=0, \quad \\
(\partial_r+\mathbf{U_K})\phi_{2} +\mathbf{U_I}\phi_{1}&=0, \quad
\end{align} 
with the same $U_{K}$ and  $U_{I}$ given above. 
For the clarity of the physics we turn on just one field $\Phi_I$ to calculate corresponding spectral function. 
 $\Phi^{I}$ is  the order parameter field that couples with spinor bilinear in the bulk.  In this paper, we will treat  it at the probe  level with  AdS background. Although the probe solution for $\Phi_{I}$ does not respect all the requirements at the horizon,   the IR region   where the probe solution blows up  by $\sim 1/r^{\Delta}$  is removed  by the presence of the horizon. Therefore it is a good approximation, unless the temperature is excessively small. 
We will separately   consider  the cases where  order parameter field 
 with  condensation  only   and  the case with source only   in order to understand  the effect of each case. 
 %

 \subsection{Discrete symmetries in AdS$_{4}$ } 
To discuss the discrete symmetry, we first list the explicit forms of  the Gamma Matrices we use. 
\bea
		&&\Gamma^t=\sigma_{1}\otimes i \sigma_{2},  
		\quad\Gamma^x=\sigma_{1}\otimes  \sigma_{1}, 
		\quad\; \;\Gamma^y=\sigma_{1}\otimes \sigma_{3}, 
		\quad\Gamma^r=\sigma_{3}\otimes \mathbb{1}, \\
	&&\Gamma^5=i\Gamma^{0123}=\sigma_{2}\otimes \mathbb{1} ,
	 \;\;\Gamma^{\mu\nu}=\frac{1}{2}\left[\Gamma^\mu, \Gamma^\nu\right], 
	\Gamma^{tx}=\mathbb{1} \otimes \sigma_{3},  
	 \;\Gamma^{ty}=\mathbb{1} \otimes -\sigma_{1}, \\
		&&\Gamma^{xy}=\mathbb{1} \otimes -i \sigma_{2}, 
		 \Gamma^{rt}=i\sigma_{2} \otimes i\sigma_{2}, 
		\;\;\Gamma^{rx}=i\sigma_{2} \otimes  \sigma_{1},
		\;\Gamma^{ry} =i\sigma_{2} \otimes \sigma_{3},\\
			&&\Gamma^{t5}=i\sigma_{3} \otimes i\sigma_{2},
			\;\Gamma^{x5}=i\sigma_{3} \otimes  \sigma_{1},
			\;\;\Gamma^{y5}=i\sigma_{3} \otimes  \sigma_{3},
			\;\Gamma^{r5}=-i\sigma_{1} \otimes  \mathbb{1}
\eea
Our convention of the tensor product is that the second factor is imbedded into each component of the first factor. 
Notice that the construction is based on $\Gamma^{\mu}=\sigma_{1} \otimes \gamma^{\mu}$, for $\mu=0,1,2$, and $\Gamma^{r}$ was chosen to satisfy the Clifford algebra $\{\Gamma^{\mu},\Gamma^{\mu}\}=2\eta^{\mu\nu}$. $\half(1\pm\Gamma^{r})$ are projections to the upper (lower) two components of the 4-component Dirac spinor. In AdS space, the bulk mass of  a field is not 
playing the role of the gap. Therefore without interaction, fermion spectrum is basically massless, and therefore  helicity is a good quantum number. The upper two components are for positive helicity while   lower two components have  negative helicity. Depending on the boundary term, some of the components are projected out. 
In this paper we will choose the upper two components of 
the first flavor and lower two of the second flavor.

 The bulk gamma matrix is $4\times 4$ and  we can decompose it  into irreducible representations of Lorentz group:    
	\be \rm
	{\bf 16}={\bf 1}(scalar)+{\bf 4}(vector)+{\bf 6}(tensor)+{\bf 4}(axial~ vector)+{\bf 1}(pseudo ~scalar), \label{class}\ee
and  we will consider each type of the interaction in detail.
  
From the boundary point of view, we have scalar and  vector interaction.  
What happened to the correspondence between the bulk and and the boundary? 
we can reclassify the the 16 AdS4 tensors in terms of 
2+1 tensors. 
\begin{itemize}
\item 4   scalars: $1, \Gamma^{5}, \Gamma^{r}, \Gamma^{r5}=\sigma^{A}\otimes  \mathbb{1}$ with $\sigma^{A}=( \mathbb{1}, \sigma^{2},\sigma^{3}, -i\sigma^{1})$ .  
\item 3 types of vectors 
 $\Gamma^{\mu}= \sigma^{1}\otimes\gamma^{\mu}$, 
 $\Gamma^{\mu 5}=i\sigma^{3}\otimes\gamma^{\mu}$, 
 $\Gamma^{r\mu}=\i\sigma^{2}\otimes\gamma^{\mu}$,
\item  3 tensors $\Gamma^{\mu\nu}=\epsilon^{\mu\nu\alpha}\mathbb{1}\otimes\gamma_{\alpha} $, where  index runs 0, 1, 2.   
\end{itemize}
We will see the similarities  in each classes. 

Below, we discuss the three discrete symmetries. ${\cal T},{\cal P},{\cal C}$ acting on the Dirac spinors and its bilinear in our gamma matrix convention. 
We need to know that the hermitian  form of interaction lagrangian is given by 
\be {\cal L}_{int} = 
\Phi_{I}{\bar\psi}_{1}\Gamma^{I}\psi_{2} + 
\Phi_{I}^{*}{\bar\psi}_{2}\Gamma^{I}\psi_{1} 
\ee
for all $\Gamma^{I}=i\mathbb{1},\Gamma^{\mu},\Gamma^{5\mu},\Gamma^{\mu\nu}$ with $\mu,\nu=t,x,y,r$.  

\begin{itemize} 
\item
The time reversal operation  is given by $ {\cal T}=T K$ where  $K$ is complex conjugation and T is a unitary matrix. From the invariance of the Dirac equation, we have  $T\Gamma^{0*}T^{-1}= -\Gamma^{0}$ and 	 $T\Gamma^{i*}T^{-1}= +\Gamma^{i}$.  Since $\Gamma^{\mu}$ ($\mu=t,x,y,r$)  are all real in our gamma matrix convention, we should have have $T=\Gamma^{1}\Gamma^{2}\Gamma^{3}$. 
Under the $\psi(t)\to\psi'(t') ={\cal T}\psi(t)=T\psi^{*}(-t)$,  
\be
{\bar\psi}_{1}\Gamma^{I}\psi_{2} \to   {\bar\psi}_{2}\Gamma^{5}\Gamma^{I\dagger}\Gamma^{5}\psi_{1} .
\ee
Therefore the invariant Hermitian  bilinears correspond  to following 8   matrices:  
\be
\Gamma^{I}= \Gamma^{5}, \Gamma^{5r}, \Gamma^{t},  \Gamma^{5i}, \Gamma^{ti}, \Gamma^{tr}. 
\ee
On the other hand, the other half with 
change sign under the time reversal operation. 
\be
\Gamma^{I}=i\mathbb{1}, \Gamma^{r},  \Gamma^{5t}, \Gamma^{i},  \Gamma^{ri}, \Gamma^{xy}.
\ee
 
\item The  parity symmetry   $(t,x,y,z)\to (t,-x,-y,-z)$ with $z=1/r$. For this, one should   imagine that two AdS  spaces  with $z>0$ and $z<0$ are patched together along the hyperplane at  $z=0$. Notice that   vierbeins are even function of $z$ because  $e^{\mu}_{a}=\delta^{\mu}_{a}\sqrt{g^{\mu\mu}}$ and the horizon of the mirror geometry is located at $- z_{H}$.    The operation $P: \psi(t,x,y,r)\to \Gamma^{0}\psi(t,-x,-y,-z)$  realizes the symmetry,    under which a fermion bilinear transforms
\be 
{\bar\psi}_{1}\Gamma^{I}\psi_{2} \to - {\bar\psi}_{1}\Gamma^{0}\Gamma^{I} \Gamma^{0}\psi_{2}.
  \ee 
Then  the invariant Hermitian quadratic forms correspond  to following 8 gamma matrices:  
\be
\Gamma^{I}=i\mathbb{1}, \Gamma^{5r},  \Gamma^{t}, \Gamma^{5i},  \Gamma^{ri}, \Gamma^{xy}.\ee
On the other hand, the other half with 
\be
\Gamma^{I}= \Gamma^{5}, \Gamma^{r}, \Gamma^{5t},  \Gamma^{i}, \Gamma^{ti}, \Gamma^{tr}. 
\ee
change sign under the parity operation. Later we will see that the fermions with interactions invariant under the Parity 
will have zero modes, that would be interpreted as a surface mode, if   there were an edge  of the boundary of the AdS.

\item The charge conjugation in our Gamma matrix convention is given by ${\cal C}=CK$ with $C=\mathbb{1}$. This is due to the reality of the $\Gamma^{a}$ with $a=t,x,y,r,5$.  Under this  symmetry,
\be 
{\bar\psi}_{1}\Gamma^{I}\psi_{2} \to {\bar\psi}_{2} \Gamma^{0}\Gamma^{I\dagger}\Gamma^{0}\psi_{1} =  {\bar\psi}_{2}\Gamma^{I}\psi_{1}. \ee 
Therefore the bilinear term is invariant if the interaction is invariant under  the  $ 1\leftrightarrow 2$ and the order parameter is real. 
\item Next,  we   define the chiral symmetry under which 
we combine the time reversal and sublattice symmetry  ${\cal S}: 1\leftrightarrow 2$,  
\be
\psi_{1}(t,x,y,r)\to \Gamma^{0}\psi^{*}_{2}(-t,x,y,r). 
\ee
It can be realized  
by ${\cal X}
=\Gamma^{0}K\cal{S}$,   so that 
\be
{\bar\psi}_{1}\Gamma^{I}\psi_{2} \to  - {\bar\psi}_{1}\Gamma^{I\dagger}\psi_{2}. 
\ee
Therefore  quadratic forms corresponding  to following 8 hermitian gamma matrices     change the sign 
\be
\Gamma^{I}= \Gamma^{5}, \Gamma^{r}, \Gamma^{5t},  \Gamma^{i}, \Gamma^{ti}, \Gamma^{tr} , 
\ee
 while  the other half    
\be
\Gamma^{I}=i\mathbb{1}, \Gamma^{5r},  \Gamma^{t}, \Gamma^{5i},  \Gamma^{ri}, \Gamma^{xy}, \ee
which are anti-hermitian matrix 
does not change sign under this symmetry operation.  
Then  the kinetic term effectively reverse the sign while the  mass term is invariant and 
  the equation of motion, hence {\it the spectrum,  is invariant as far as the order parameter is real and the $\Gamma^{I}$ is hermitian}. 
Notice  the set of spectral symmetry of ${\cal X}$ is precisely complements of that of the parity symmetry. 
Notice also that   $\cal X$ could be a possible symmetry of the system   because the bulk  mass terms were chosen as 
$-im({\bar \psi}_{1}\psi_{1}-{\bar \psi}_{2}\psi_{2} ) $ instead of 
$-im({\bar \psi}_{1}\psi_{1}+{\bar \psi}_{2}\psi_{2} ) $, which explains our  choice of the opposite  signs in mass terms.  
However, such a change of mass term is a unitaty operation and can not change the spectrum. On the other hand 
{The parity is a symmetry regardless of such sign choices.} 
As we will see, the spectrum of our theory follows $\cal P$ while  its dual   follows $\cal X$. 
\end{itemize}


 \section{Classifying  the  spectrum by the order parameter for 2 flavours}
 we classify the spectrum  into scalar, vector, and tensor  along the line we discussed above. In all the figure below, we should keep in mind that the $k_{x}$ and the vertical one represents 
 either $\omega$ mostly except the fixed $\omega$ slice, where the vertical axis is  $k_{y}$.

\subsection{Summary of spectral  features }
Here,  we   classify, summarize and tabulate  some essential spectral features. 
\begin{description}
\item[Spectral Classification]  Following the discussion below \ref{class}, we classify the spectrum according to the 2+1   Lorentz tensor. 
\\$\bullet $ There are 4 scalars.   $\mathbb{1}, \Gamma^{5},\Gamma^{r},\Gamma^{5r}$.  The first two  were described above.   For the gauge invariant fields $B_{\mu}$, we should set $B_{r}=B_{5r}=0$. In fact, even for non-gauge invariant case,the last two  are identical to the zero   Yukawa coupling  in our gamma matrix representation. 
For scalar interactions, the roles of source and condensation are qualitatively the same. 
 \\
$\bullet $  There are three classes of vectors:
$B_{\mu}, B_{r\mu}$ and $B_{5\mu}$.
The source creates the split cones and the condensation creates just asymmetry. 
The first two are invariant under the parity  symmetry showing  zero mode related features like  Fermi-arc and surface states(Ribbon band).
\\
$\bullet $ There are 3 rank 2-tensor terms: $\Gamma^{xy}, \Gamma^{tx},\Gamma^{ty}, $. The first one is parity  invariant and   has zero modes.  

\item [Gap vs zero modes with scalar order] Out of the 16 interaction types, only parity   symmetry breaking  scalar  interaction  with $\Gamma=\Gamma^{5}$) 
 creates a gap without ambiguity. Both source and the condensation create gaps.  \\
 On the other hand,  parity invariant scalar  with $\Gamma=i\mathbb{1}$ has a zero mode Dirac cone in spectrum, which is much sharper than the case of the non-interacting case due to the transfer of the spectral weight  to the zero mode by  the interaction. The genuine  physical system with full gap will be described by this coupling. 
 
\item[Pseudo scalar] 
When the interaction is parity non-invariant, 
the spectrum has pseudo gap apart from $\Gamma^{5}$ which produces real gap. Seven  interactions corresponding to 
$$
\Gamma^{I}= \Gamma^{r}, \Gamma^{5t},  \Gamma^{i}, \Gamma^{ti}, \Gamma^{tr}. 
$$
have pseudo gaps. Therefore the pseudo gap is a typical phenomena   rather than an exception for general interaction in this theory, while the true gap is a rare phenomena. 
\item [Fermi-arc] For vectors $B_{\mu}$, $B_{5\mu}$ and $B_{r\mu}$ the role of source term is to generate the split Dirac cones along $k_{\mu}$ direction, $\mu=0,1,2$,   while that of the condensation is to generate an anisotropy. This is  just like the flat space cases. 
However, there is a very interesting phenomena when the interaction term is  invariant under the parity:  that is,  for $B^{5}_{i},B_{ri}$ $i=x,y$,
there exist a spectral line connecting the tips of two Dirac cones at $\omega=0$ plane. This   resembles   
the ``Fermi-arc'' in the study of Dirac   or Weyl-semi metal.  
At the plane  $\omega=\epsilon \neq 0$ there is also a spectral line(s) connecting the surfaces of the cones. This is precisely the same as the surface modes of topological materials \cite{Armitage:2017cjs}. 
 If we collect all the slice of different $\epsilon$,  the   surface modes form a Ribbon shaped band. See Figure \ref{Ribbon}.
 One should notice that we did not introduce the edge of 2+1 dimensional system. In our case, the ``surface states''  is properties of the bulk zero modes which does not depends on the presence of the edge. Perhaps this is general phenomena 
 and responsible to the bulk-edge correspondence. 
 \footnote{This is analogous to the Faraday's law 
 $\oint_{C} {\bf dl}\cdot{\bf E} =-\frac{\partial {\bf B}}{\partial t} $
 where  left hand side  is non-zero regardless of the presence of the real circuit along the curve,  if there is a time dependent magnetic flux.} 
 
 \begin{figure}[ht!]
			\centering
        {\includegraphics[width=6cm]{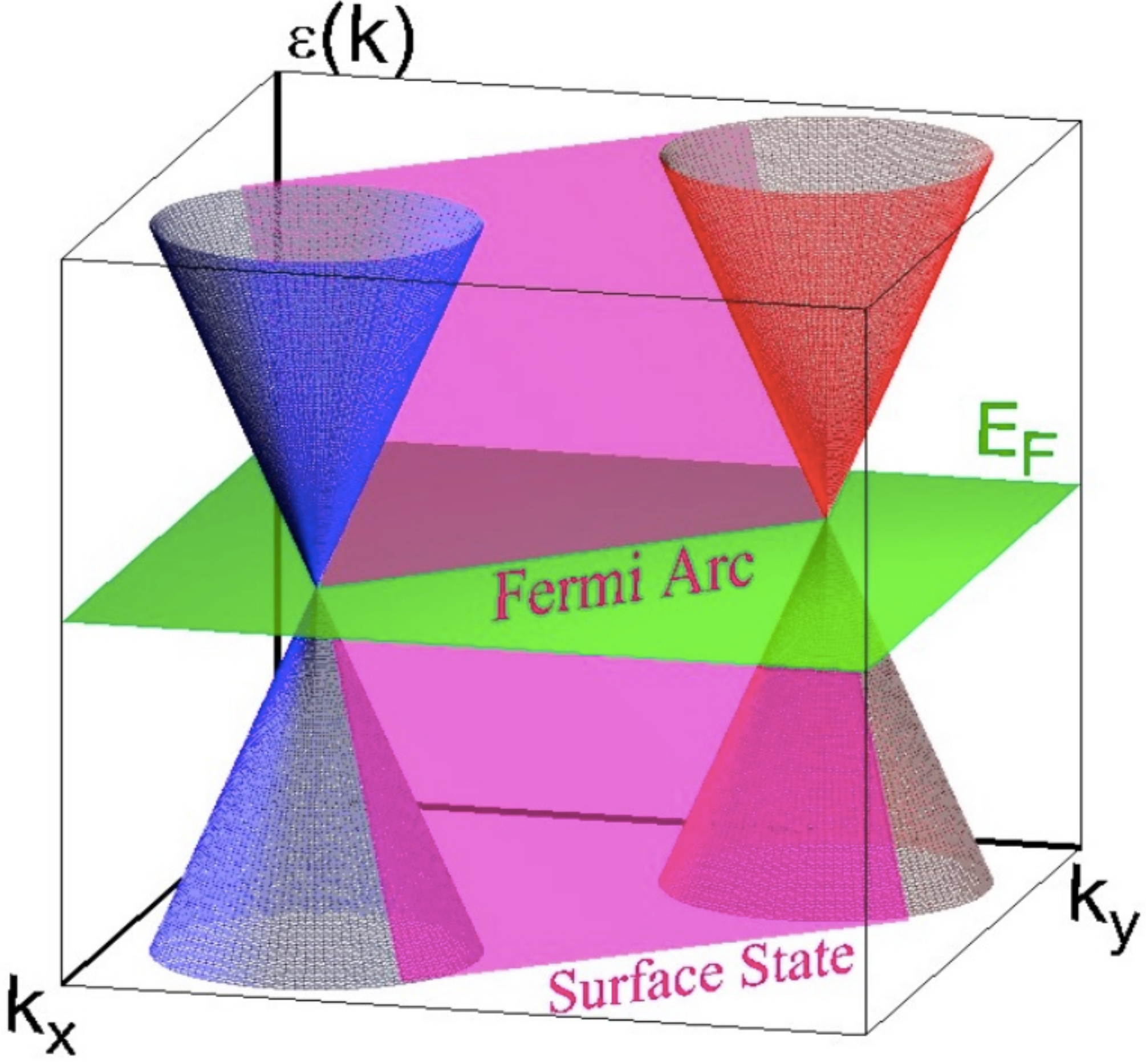}}			
			\caption{Splitted Dirac Cones, Fermi-arc  and Surface modes. Our analysis indicates that the so called ``surface states'' are bulk zero modes which exists regardless of the presence of the edge of the physical system. The figure came from   \cite{Armitage:2017cjs}. 
			}
			   \label{Ribbon}
		\end{figure} 
\item [Flat band] $B_{xy}$ interaction introduces a flat band which is a disk like isolated band at the fermi level $\omega=0$. If  chemical  potential is applied, the disk bend like a bowl and the fermi level shifts.  

\item [Zero mode and parity  symmetry] 
In the presence of the background field $B_{I}$ with coupling 
$B_{I}{\bar\psi}\Gamma^{I}\psi$, 
the spectrum shows the zero modes if the quadratic form is   parity  invariant. 
\item[Duality] If we change the boundary term to 
$ S_{ bdry}=\half \int_{bdry} d^{3}x\; i({\bar\psi}_{1}\psi_{1}-{\bar\psi}_{2}\psi_{2})\label{bdryaction},$
then the spectrum of dual pairs are exchanged. 
By the dual pair, we mean one of following set of pairs: 
$$(\Phi, \Phi_{5}), (B_{\mu},B_{5\mu}), (B_{\mu\nu}, \epsilon_{\mu\nu\alpha\beta}B^{\alpha\beta}),  $$
with   indices running $t,x,y,r$. 
We found that,  in this case, the presence of the zero mode are protected by the chiral symmetry $\cal X$ we defined earlier.  
\item[Order without rotational symmetry breaking]
The presence of the non-vanishing order parameter  field means the  breaking of the some rotational or Lorentz symmetry from the bulk point of view.  However,  from the boundary point of view,  some  order parametrs involving $r$-index, like $B_{rt}$ does not have obvious symmetry breaking interpretation and therefore they  can be interpreted as   `orders  without symmetry breaking'. 
\end{description}

\begin{table}[hbt!]
	\centering
	\begin{tabular}{|c|c|c|c|c|c|}
		\hline
		\multicolumn{2}{|c|}{Order p./ Fig.\#} & Gap & zero mode & spectral   feature  & possible dual system \\
		\hline \hline
		\multirow{2}{*}{$\Phi_{5}$} & s/4(a)  & $\ocircle$    & \multirow{2}{*}{$\times$}          & \multirow{2}{*}{ gap }          &   {  RS(real $\Phi$)} \\ \cline{2-3} 
		& c/4(b)  & $\ocircle$    &          &          & SC(complex $\Phi$) \\
		\hline
		\multirow{2}{*}{$i\Phi$} & s/4(c)  & $\times$    & \multirow{2}{*}{$\ocircle$}          & \multirow{2}{*}{ Dirac cone }          & \multirow{2}{*}{ Majorana Fermion in SC} \\
		 \cline{2-3}
		& c/4(d)   & $\times$    &           &           &\\
		\hline
		\multirow{2}{*}{$B^{5}_{r}$}         & s/4e         & $\times$    & \multirow{2}{*}{$\ocircle$}          & \multirow{2}{*}{Non-coupling}          &  \multirow{2}{*}{NA } \\ \cline{2-3}
		& c/4e         & $\times$    &           &           &\\
		\hline
		\multirow{2}{*}{$B_{r}$}         & s/4e         & $\times$    & \multirow{2}{*}{$\times$}          & \multirow{2}{*}{Non-coupling}          & \multirow{2}{*}{NA}\\ \cline{2-3}
		& c/4e         & $\times$    &           &          &\\
		\hline\hline			
		\multirow{2}{*}{$B^{5}_{i}$}         & s/6abc         & $\times$    & \multirow{2}{*}{$\ocircle$}          & {Split cones  }          &  \multirow{2}{*}{ Top. semi-metal} \\ 
		\cline{2-3}
		& c/6def           & $\times$    &           &      Fermi arc    &\\
		\hline
\multirow{2}{*}{$B_{i}$}         & s/6ghi         & $\times$    & \multirow{2}{*}{$\times$}          & {Split cones}          & \multirow{2}{*}{ NA}\\ \cline{2-3}
		& c/6jkl         & $\times$    &           &   pseudo gap      &\\
		\hline
		\multirow{2}{*}{$B^{5}_{t}$}         & s/7abc     & $\times$    & \multirow{2}{*}{$\times$}          & \multirow{2}{*}{Rot. Sym }          & \multirow{2}{*}{NA }\\ \cline{2-3}
		& c/7ghi             & $\times$    &           &           &\\
		\hline
		\multirow{2}{*}{$B_{t}$}         & s/7def        & $\times$    & \multirow{2}{*}{$\ocircle$}          & \multirow{2}{*}{Nodal line}          & \multirow{2}{*}{Top. semi-metal }\\ \cline{2-3}
		& c/7jkl       & $\times$    &           &           &\\
		\hline
		\multirow{2}{*}{$B_{rt}$}         & s/8c         & $\times$    & \multirow{2}{*}{$\times$}          & \multirow{2}{*}{Marginal gap}          & \multirow{2}{*}{NA}\\ \cline{2-3}
		& c/8d         & $\times$    &           &           &\\
		\hline
		\multirow{2}{*}{$B_{ri}$}         & s/9ghi        & $\times$    & \multirow{2}{*}{$\ocircle$}          & {Split cones}          & \multirow{2}{*}{Top. Ins.}\\ \cline{2-3}
		& c/9jkl        & $\times$    &           &    Fermi-arc       &\\
		\hline\hline
				\multirow{2}{*}{$B_{xy}$} & s/8a  & $\triangle$    & \multirow{2}{*}{$\ocircle$}          & \multirow{2}{*}{Disk flat band }          & {twisted bi-layer graphene }\\ \cline{2-3}
		& c/8b & $\triangle$    &           &           &Kondo lattice\\
\hline		
\multirow{2}{*}{$B_{ti}$}         & s/9abc        & $\times$    & \multirow{2}{*}{$\times$}          & \multirow{2}{*}{Split cones, Fermi-arc}          & \multirow{2}{*}{Top. Ins.}\\ \cline{2-3}
		& c/9def        & $\times$    &           &           &\\
		\hline				
	\end{tabular}
	\label{tab:class2}
	\caption{In the table of "Gap", $\ocircle$ denotes gap at the fermi-level, $\triangle$ represents  gap off the fermi level and $\times$ is gapless. SC=superconductivity, RS=Random Singlet. 
	$A(k_x,\omega)$ means we consider the spectral function $A$ as the function of $k_x$ and $\omega$. Under $k_{x}\leftrightarrow k_{y}$ those with one spatial  index are assymetric. All others are symmetric. 
	NA=not available.
 }
\end{table}

\begin{figure}[ht!]
	\centering
	\subfigure[ $B_{xy,c}=0$]
	{\includegraphics[width=3.5cm]{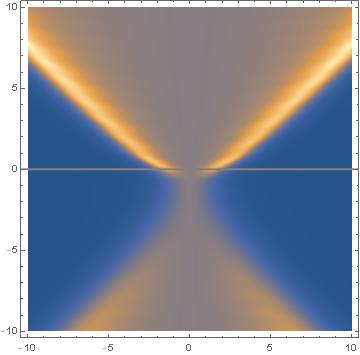}}
\subfigure[ $B_{xy,c}=5$]
	{\includegraphics[width=3.5cm]{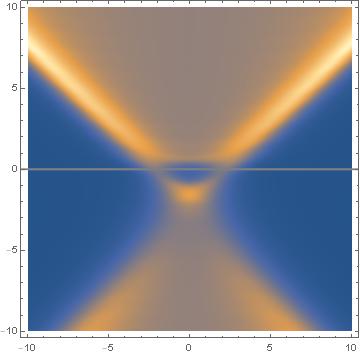}}
\subfigure[ $B_{xy,c}=10$]
	{\includegraphics[width=3.5cm]{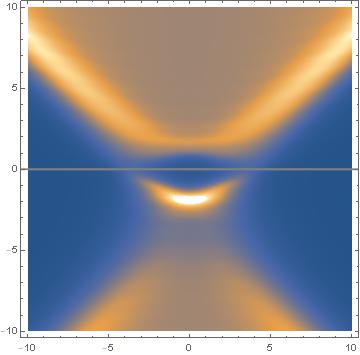}}
	\subfigure[ Kondo lattice]
	{\includegraphics[width=4.5cm]{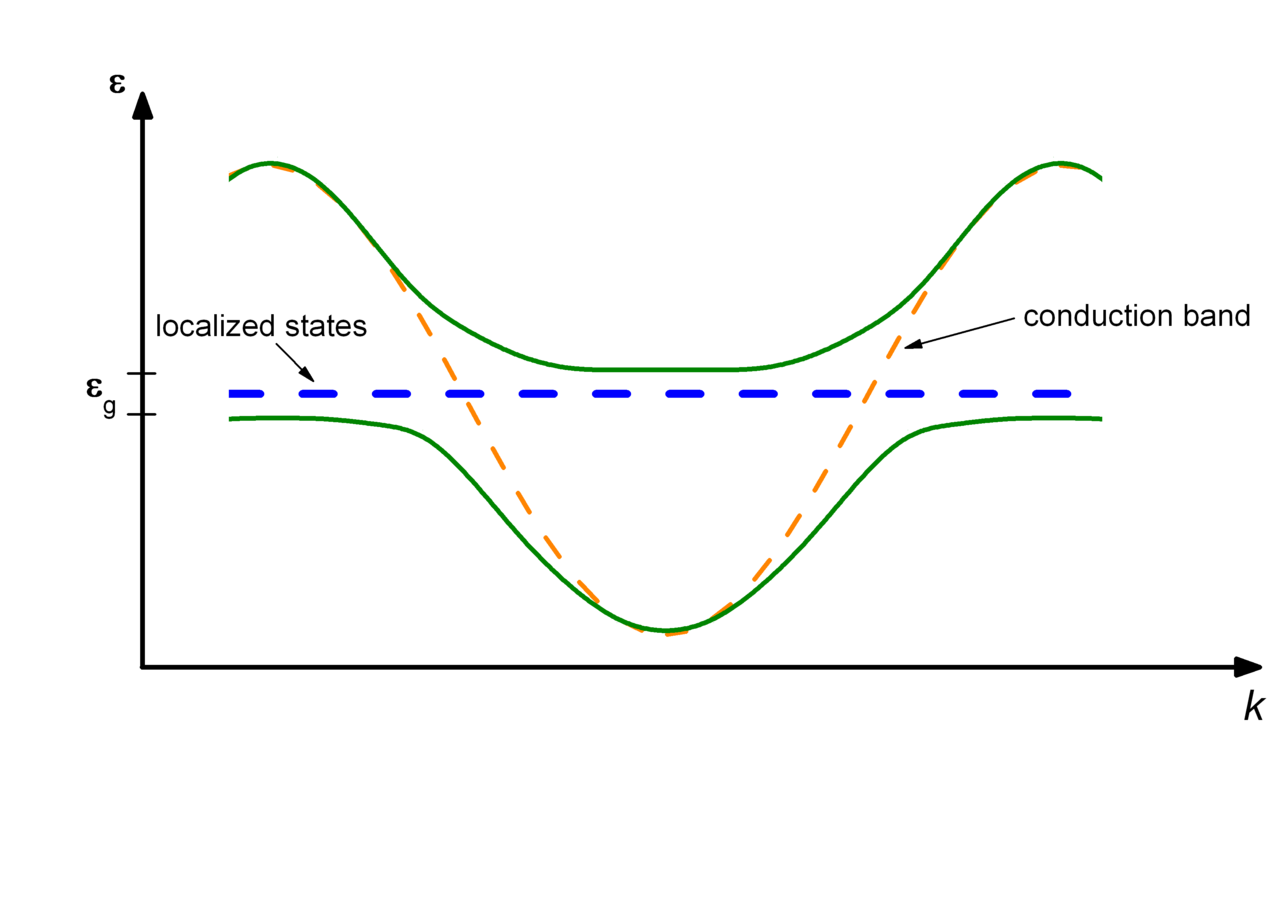}}	
 	\caption{(a-c) Formation process of bent flat band by as we change the stregnth of the coupling. 	From the left to right $B_{xyc}=0,5,10$. The chemical potential is fixed to be $\mu=2\sqrt3$.  (d) Formation of flat band by hybridization of localized state and conducting state. 
}
	\label{fig:kondo} 
\end{figure}

The table 1 summarizes all the features we found.
 We attributed the presence of the zero modes to the protection of the  parity  invariance. The zero mode  is of course  the key for the surface states.  The    presence of the zero mode results in  the bright crossing of the Dirac cone with the Fermi-level. This means that the zero modes create sharp Fermi-surface, which was orginally fuzzy due to the strong interaction at the boundary. This is one of the most interesting observation made in this paper.   That is,  the parity invariant interaction can make a strongly interacting system be 
 fermi-liquid like.  Below  in the figure \ref{fig:kondo} we give comparison of the spectrum with coupling  $B_{xy}{\bar \psi}\Gamma^{xy}\psi$ 
in the presence of chemical potential and that of the heavy fermion in Kondo lattice. 
More explicit comparison with the experimental data is left as a future project. 

 \subsection{ Spectral Function (SF) with   Scalar interaction}
 \subsubsection{ parity  symmetry  breaking   case :
	 ${\cal L}_{int}= \Phi_{5}({\bar\psi}_{1}\Gamma^{5}\psi_{2}+{\bar\psi}_{2}\Gamma^{5}\psi_{1})$ }
\begin{figure}[ht!]
	\centering
	\subfigure[ $\Phi_{5}$, s]
	{\includegraphics[width=3cm]{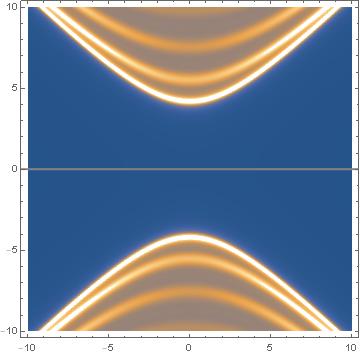}}
	\subfigure[ $\Phi_{5}$, c]
	{\includegraphics[width=3cm]{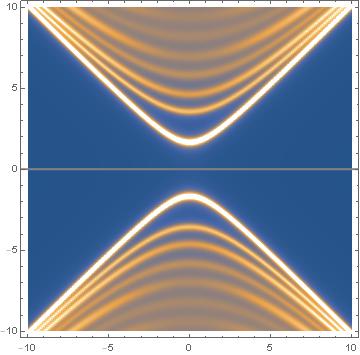}}
	\subfigure[$\Phi$, s]
	{\includegraphics[width=3cm]{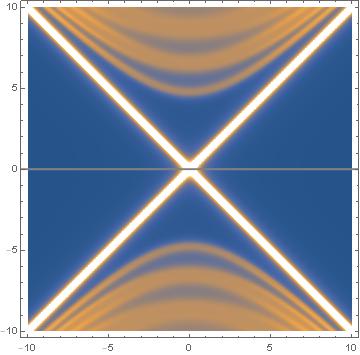}}
	\subfigure[ $\Phi$. s]
	{\includegraphics[width=3cm]{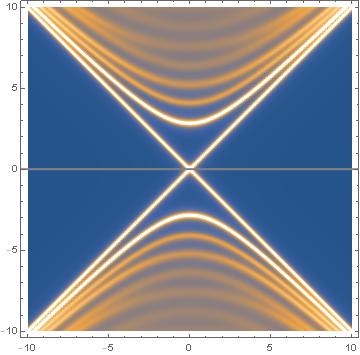}}	
		\subfigure[ $B_{r},B_{5r}$, s, c]
	{\includegraphics[width=3cm]{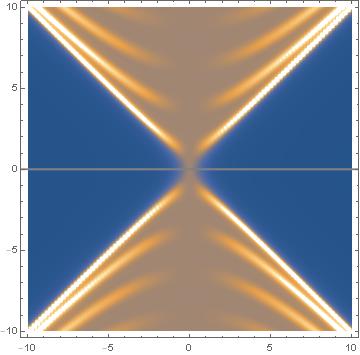}}	
	\caption{Spectral Function(SF) (a,b) with parity
	 breaking scalar. (a) with source only.  Gap $\Delta\sim M_{05}$.    (b)  with condensation only,   $\Delta\sim \sqrt{M_{5}}$;
	 (c,d) with parity invariant  scalar. Notice the zero modes.   (c) source only (d)  condensation only. (e) $B_{r},B_{rt}$ shows the   spectrum  of zero coupling due to the gamma matrix structrure. }
	\label{fig:scalar} 
\end{figure}

We begin with the   simplest case   where the order parameter field is scalar field. We choose $m_{\Phi}^2=-2$ in $AdS_4$ for simplicity. Then \cite{erlich2005qcd,oh2019holographic}
\be
\Phi_{5}=M_{05} z + M_{5} z^2,\ee
 in the probe limit.   We consider source only and  condensation only cases separately. 
 	\paragraph{ Scalar Source: $M_{05}$   }
	The scalar source is usually interpreted as a mass of the boundary fermion. Indeed our result given in   
	 the Figure \ref{fig:scalar}(a), where  we  draw the   spectral function (SF)	in the presence of    scalar   with  source term only,  fulfill such expectation.  
 
 \paragraph{ Scalar Condensation: $M_{5}$. }
This case describes the spontaneous scalar  condensation. For complex $\Phi$ with nonzero $M$  it describe the cooper pair condensation  while for real   case it may describe a chiral condensation or a random spin singlet condensation where lattice spins  pair up to form singlets,  the dimers,  in random direction so that there is no net magnetic ordering.   In fact, in lattice models with  antiferromagnetic coupling, the ground state is anti-ferromagneticaly  ordered   if  frustrations  and randomness are small enough. 
On the other hand,  it has random singlet (RS) state\cite{bhatt1982scaling,paalanen1988thermodynamic,PhysRevLett.48.344, guo1994quantum}   if there   is a randomness, a  distribution of  next-nearest site couplings.  Whether a RS like state has a gap or not depends on the   details of the lattice symmetry  as well as the size of the randomness \cite{PhysRevX.8.031028,PhysRevB.98.054422,Uematsu_2018,Liu_2018,PhysRevLett.123.087201,Kawamura_2019,Im}. 
Our philosophy is to bypass all such details and characterize the system only by a few order parameter, assuming this is possible at least near the critical points.  
From our calculation,  a RS state with gap is described by a scalar order. 
Notice that the dipole type interaction $F_{rt}{\bar\psi}\gamma^{rt}\psi$,  which was used to study the Mott physics \cite{Edalati:2010ww,Edalati:2010ge,Seo:2018hrc}, does not generate a true gap,  because its density of state does not really has a gap although its spectral function has gap like features in small momentum region.  
This is   because the spectral function shows a band that approaches to the fermi level for large momentum. 

 \paragraph{ Spectrum in  potential picture }
 More characteristic feature is the  appearance of  Kaluza-Klein (KK) modes in  the Figure   \ref{fig:scalar},  	which is due to the effective $z^{2}$ Schr\"odinger potential  for large $z$ generated by the condensation part: the effective Schr\"odinger potential 
$V\sim \Phi_{5}^{2}/z^{2}$ which goes like $\sim M_{5}^{2}z^{2}$ for large $z$ 
 \cite{oh2019holographic}.  	   
Comparing the effect of the scalar condensation with that of the scalar source, 	the   	 gap is generated by  condensation is smaller than that  generated by the source, as shown in \ref{fig:scalar} (a,b).

\begin{figure}[ht!]
	\centering
	{\includegraphics[width=5cm]{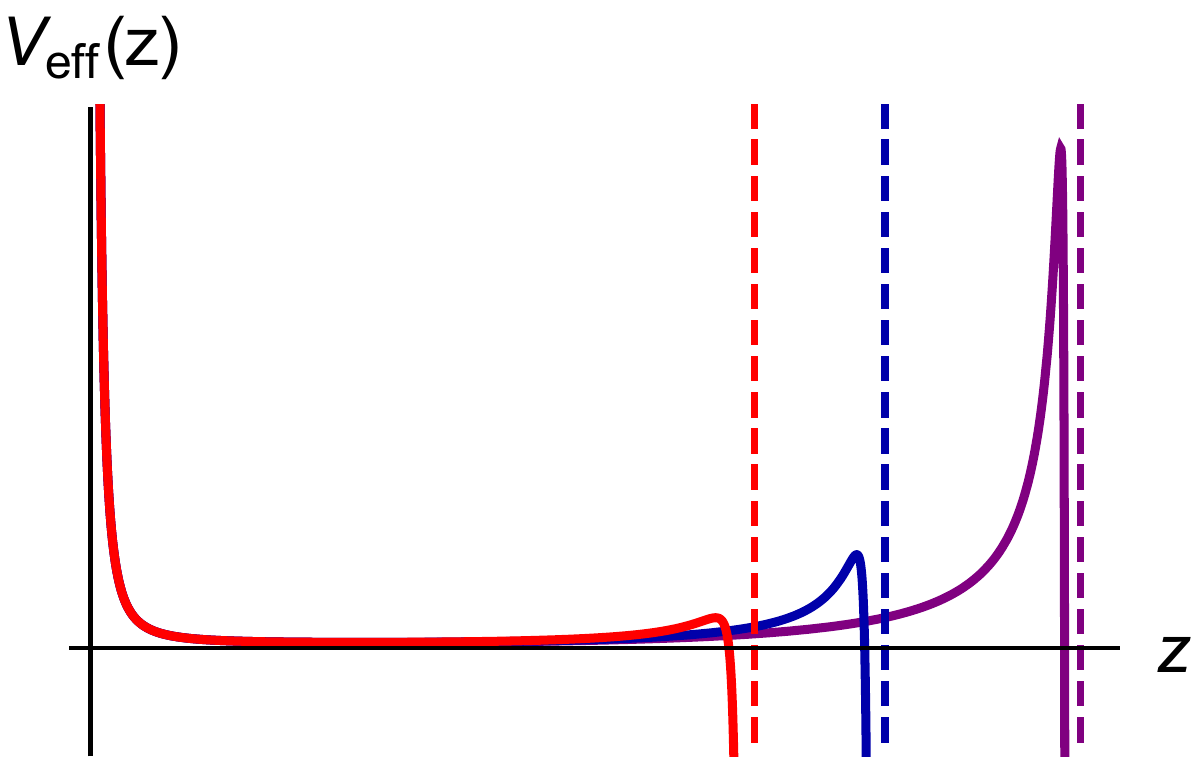}}
	\caption{ Shape of potential near the horizon. 		Dashed lines are the event horizons at a few temperatures.  As T increases,  horizon moves out in $z$-coordinate.}   \label{fig:esp}
\end{figure}

In the presence of chemical potential or temperature,  the effect  of  $z^2$ term is suppressed  because both $T$ and $ \mu$ increase the  horizon  size $r_{0}$ and the region  'inside' the black hole, $z>z_{0} =1/r_{0}$ is cut out. Then  the rising potential $z^2$ also disappear, and the potential near the horizon collapses into $-\infty$ because  near the horizon,
\begin{align}
V_{eff}(z_H)\sim - \frac{4+w^2 {z_{H}}^2}{16(z-z_H)^2},
\end{align}  
Furthermore the solution should satisfy the infalling boundary condition, so that 
instead of the infinitely many    clean quantized eigenvalues (KK modes),  only finitely many imaginary eigenvalues due to the tunneling to the horizon   appears.
 See figure \ref{fig:esp}. 
 This explains the fuzziness and disappearance of KK modes in \ref{fig:scalar}(d) in the presence of the chemical potential.
For the vector and tensor cases, there can be a pole between the horizon and the boundary.

  We emphasize that this case is not related to the rotational symmetry breaking. The $Z_{2}$ symmetry is not encoded in this model either. 
  So one natural candidate is the spin liquid with a gap\cite{Fu_2015,Oh:2018wfn}. 
  This case may  be also useful to describe the coupling between the  localized (lattice)    spin net work  and the  itinerant electron, namely the Kondo physics. 
  
 \subsubsection{ Parity preserving  Scalar interaction :
 	 ${\cal L}_{int}=i \Phi({\bar\psi}_{1} \psi_{2}+{\bar\psi}_{2}  \psi_{1})$ }
	For the consistency with scalar model, we choose the $m_{\Phi}^2=-2$ so that we still have $\Phi=M_{0} z + M z^2$. The spectrum is gapped for both source and condensation. 
	For the latter case zero chemical potential case shows sharp KK modes.  Compared with  the scalar case,   spectrum is sharper.  
See Figure \ref{fig:scalar} (c,d) 
 for the spectral functions with pseudo scalar source and  condensation respectively.  The system  is similar to the scalar but   with Parity symmetry broken. The most famous case is the pion condensation in nuclear physics.  

\subsection{Vectors }
From the 2+1 dimension boundary point of view there are 
three classes  of vectors:  
$B_{\mu}, B_{r\mu}$ and $B_{5\mu}$, respectively. 
In each class, 
the source shifts the two degenerate Dirac cones, one to  negative and the other to positive  $k_{\mu}$ directions.  
The last two classes are invariant under the parity showing  zero mode related feature like  Fermi-arc and surface states(Ribbon band). See figures   \ref{fig:Bx}, and  \ref{fig:Bx2_slice}.
	\subsubsection{Polar Vector:   ${\cal L}_{int}=   \boldmath{iB_{\mu}} ({\bar\psi}_{1}\Gamma^{\mu}\psi_{2}-{\bar\psi}_{2}\Gamma^{\mu}\psi_{1})$ }
   $B_\mu$ is the extension of the source field that couples to boundary fermion current   ${\bar \chi }\Gamma^\mu\chi$. 
 We can set the mass of vector and axial-vector order field  to zero. Then, 
	$B_{i}=B^{(0)}_{i}+B^{(1)}_{i} z^{1} $. 
 Notice that there are asymmetry between $x$ and $y$ direction. 
Therefore when $B_{x} \rightarrow B_{y}$,  it has to be followed by $k_x\to k_y$ at the same time.
\paragraph{Source} SF for  $B_{x}$  coupling with source only  is just superposition of two shifted SF's of  zero coupling cases along the $k_{x}$.  Different flavors shift in opposite directions.
 \paragraph{Condensation}:  
		Anisotropy instead of cone splitting is created. 
	See Figure and  \ref{fig:Bx2_slice} and  \ref{fig:Bx} . 

\subsubsection{pseudo Vector:   ${\cal L}_{int}=   \boldmath{iB_{5\mu}} ({\bar\psi}_{1}\Gamma^{5\mu}\psi_{2}-{\bar\psi}_{2}\Gamma^{5\mu}\psi_{1})$ }
Pseudo vectors mostly follows the pattern of polar vectors. One important point to emphasize is the line shaped zero mode at $\omega=0$ plane.  At higher slice $\omega=2$, there are also lines connecting two cicles representing the two shifted Dirac cones. See figure \ref{fig:Bx2_slice}(b)(c).  So the total 3 dimensional figure is like Figure \ref{Ribbon} with double of the surface modes, which we call Ribbon bands. 

\subsubsection{radial Vector:   ${\cal L}_{int}=   \boldmath{iB_{r\mu}} ({\bar\psi}_{1}\Gamma^{r\mu}\psi_{2}-{\bar\psi}_{2}\Gamma^{r\mu}\psi_{1})$ }
\begin{figure}[ht!]
 	\centering	  
	\subfigure[$B_{x}$ s]
	{\includegraphics[width=2.5cm]{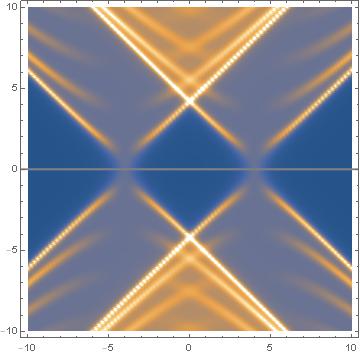}}
	\subfigure[$\omega=0$, $B_{x}$ s]
	{\includegraphics[width=2.5cm]{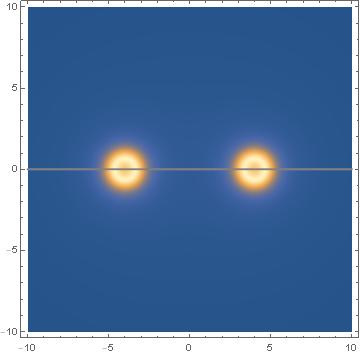}}
	\subfigure[$\omega=2$, $B_{x}$ s]
	{\includegraphics[width=2.5cm]{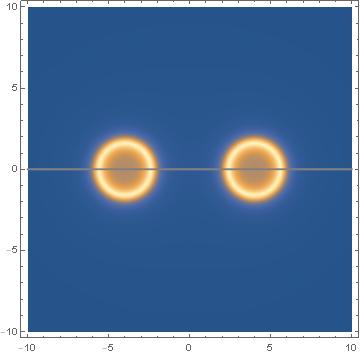}}
	\subfigure[$B_{x}$ c]
	{\includegraphics[width=2.5cm]{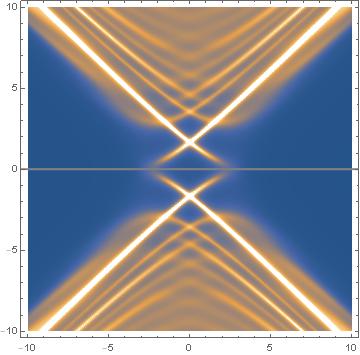}}
	\subfigure[$\omega=0$, $B_{x}$ c]
	{\includegraphics[width=2.5cm]{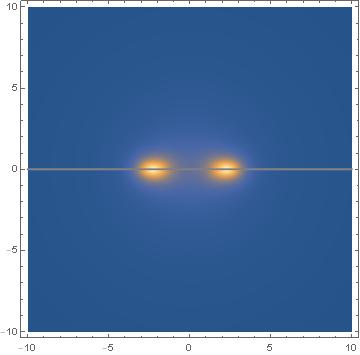}}
	\subfigure[$\omega=2$, $B_{x}$ c]
	{\includegraphics[width=2.5cm]{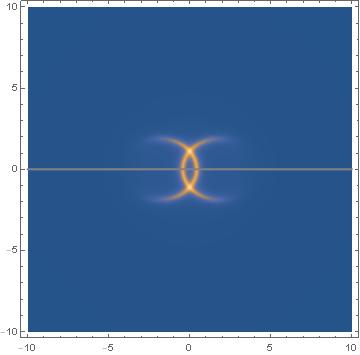}}\\
		\subfigure[$B_{5x}$ s]
	{\includegraphics[width=2.5cm]{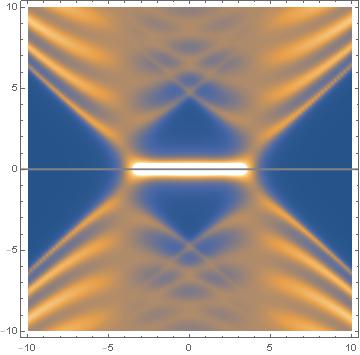}}
	\subfigure[$\omega=0$, $B_{5x}$ s]
	{\includegraphics[width=2.5cm]{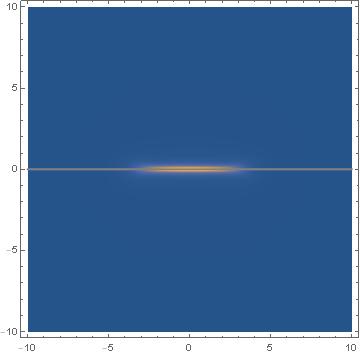}}
	\subfigure[$\omega=2$, $B_{5x}$ s]
	{\includegraphics[width=2.5cm]{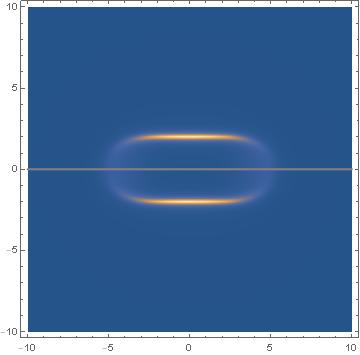}}
	\subfigure[$B_{5x}$ c]
	{\includegraphics[width=2.5cm]{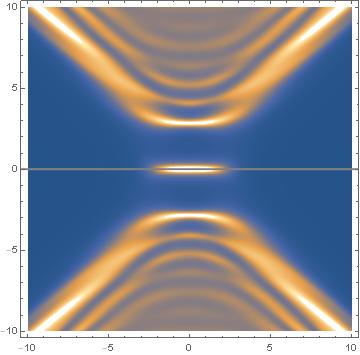}}
	\subfigure[$\omega=0$, $B_{5x}$ c]
	{\includegraphics[width=2.5cm]{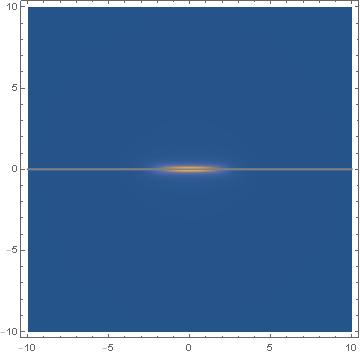}}
	\subfigure[$\omega=2$, $B_{5x}$ c]
	{\includegraphics[width=2.5cm]{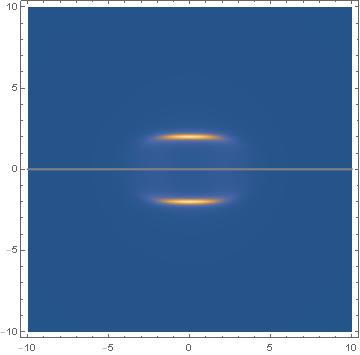}}
	\caption{(adgj)Spectral Functions of $B_{x}$ and $B_{5x}$ 
	 with additional sliced view in  $k_{x},k_{y}$ plane of $\omega=0,2$ slices.  Notice that source splits the degenerated Dirac cones. $B_{x}$ has zero modes nut   $B_{5x}$ does not. In all figures, we used $B_{*} =4 $.
	}   \label{fig:Bx2_slice}
\end{figure}

\begin{figure}[ht!]
	\centering	
	\subfigure[s, $B_{5t(0)}$]
	{\includegraphics[width=2.5cm]{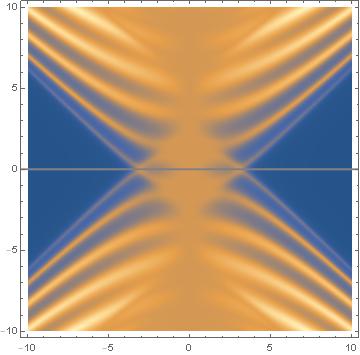}}
	\subfigure[$\omega=0$, $B_{5t} s$]
		{\includegraphics[width=2.5cm]{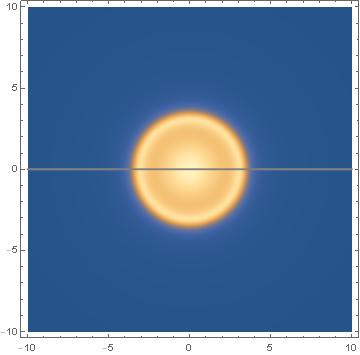}}
	\subfigure[s, $B_{5y(0)}$]
	{\includegraphics[width=2.5cm]{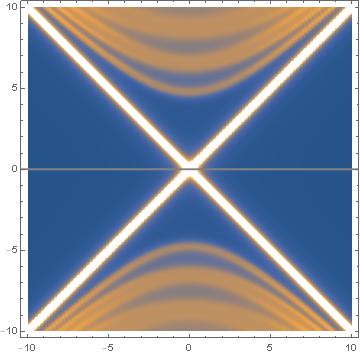}}
	\subfigure[s, $B_{t(0)}$]
	{\includegraphics[width=2.5cm]{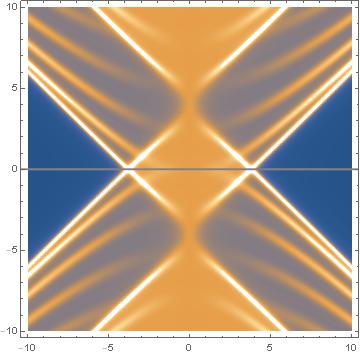}}
	\subfigure[$\omega=0$, $B_{t} s$]
{\includegraphics[width=2.5cm]{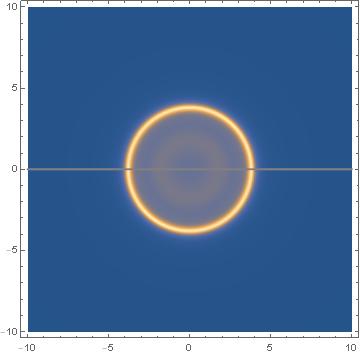}}
	\subfigure[s, $B_{y(0)}$]
	{\includegraphics[width=2.5cm]{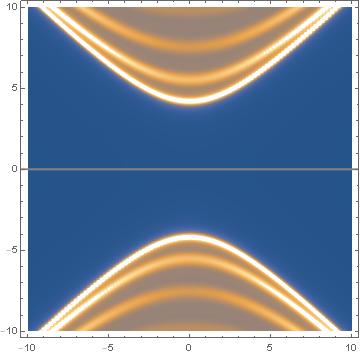}}
	\\
	\subfigure[c, $B_{5t(0)}$]
	{\includegraphics[width=2.5cm]{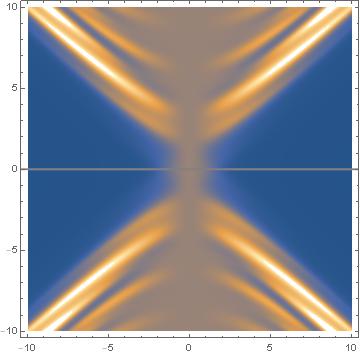}}
	\subfigure[$\omega=0$, $B_{5t} c$]
	{\includegraphics[width=2.5cm]{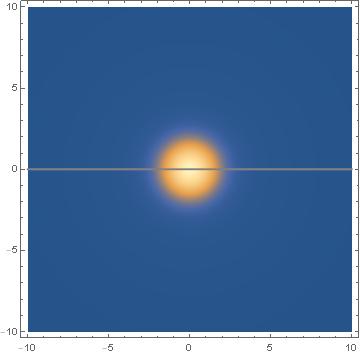}}	
	\subfigure[c, $B_{5y(0)}$]
	{\includegraphics[width=2.5cm]{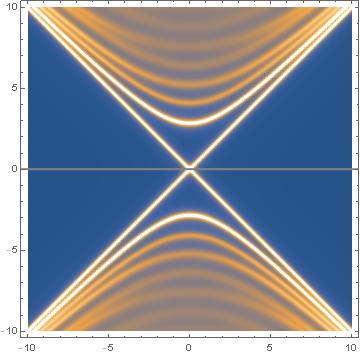}}
	\subfigure[c, $B_{t(0)}$]
	{\includegraphics[width=2.5cm]{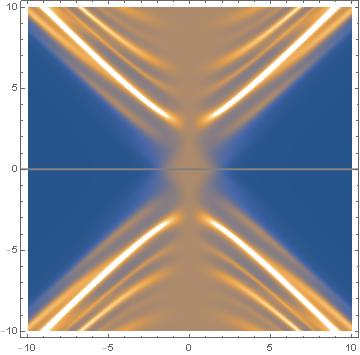}}
	\subfigure[$\omega=0$, $B_{t} c$]
	{\includegraphics[width=2.5cm]{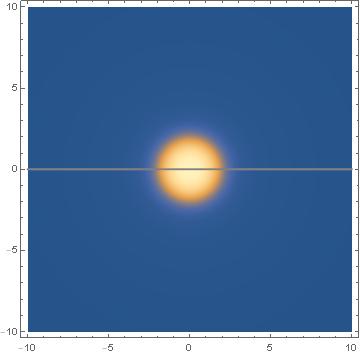}}	
	\subfigure[c, $B_{y(0)}$]
	{\includegraphics[width=2.5cm]{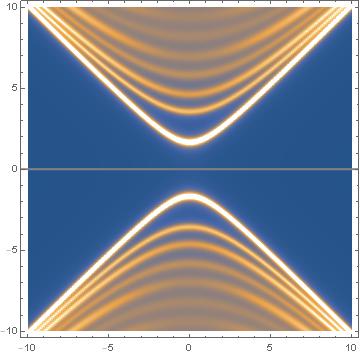}}
	\caption{Spectral function  with  (pseudo) vector source interactions (a-f), and  SF with  (pseudo)-vector  
	condensation (g-l).  s means source and c means condensation. 
}   \label{fig:Bx}
\end{figure}

Radial vectors follows the pattern of polar vectors   including zero modes. See Figure \ref{fig:Bx} and  \ref{fig:Bx2_slice}.

\newpage
\subsection{Antisymmetric 2-Tensor }
6 anti-symmetric rank 2 tensors can be decomposed into 
three $B_{\mu r}, \mu=0,1,2$ and the rests $B_{tx}, B_{ty},B_{xy}$.    The former was already described above. 

\begin{figure}[ht!]
	\centering
	\subfigure[$B_{xy(-1)}$ s]
	{\includegraphics[width=2.5cm]{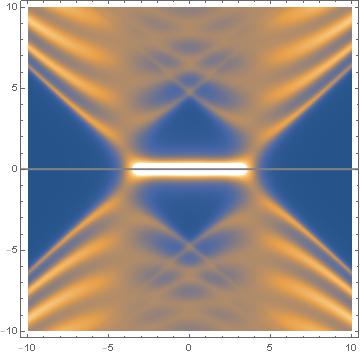}}
		\subfigure[$B_{xy(-1)}$ s]
{\includegraphics[width=2.5cm]{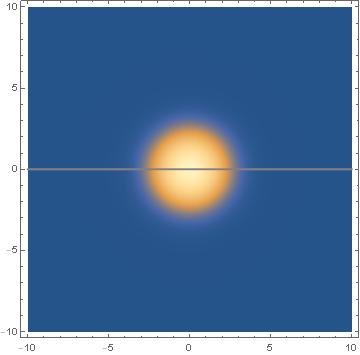}}
	\subfigure[$B_{xy(0)}$ c]
	{\includegraphics[width=2.5cm]{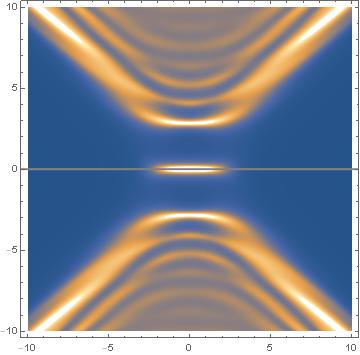}}
	\subfigure[ $B_{tr(-1)}$ s]
	{\includegraphics[width=2.5cm]{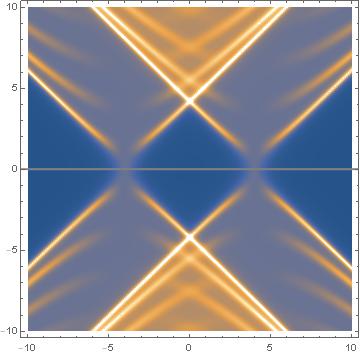}} 
	\subfigure[$B_{tr(0)}$ c]
	{\includegraphics[width=2.5cm]{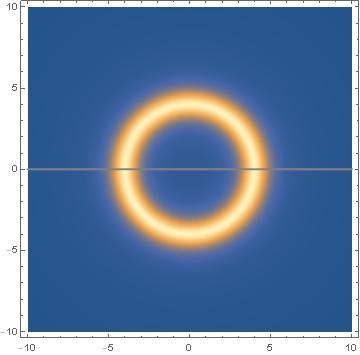}}
	\subfigure[$B_{tr(0)}$ c]
	{\includegraphics[width=2.5cm]{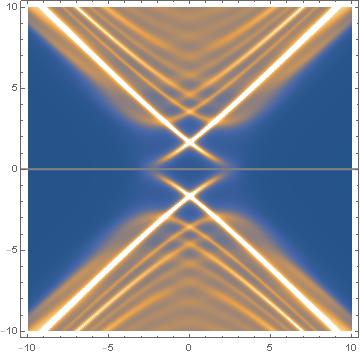}} 
\caption{Spectral function  with various types  tensor interaction, which is decomposed into 2+1 radial vector $B_{\mu r}$'s (ab) and 2-tensor $B_{xy}$  (cd).   Notice the     zero mode Disk in $B_{xy}$. There are rotational symmetry in $B_{tr},B_{xy}$.  }   
	\label{fig:Bdd2_s}
\end{figure}

\begin{figure}[ht!]
	\centering	
	\subfigure[$B_{ty(0)}$ s]
	{\includegraphics[width=2.5cm]{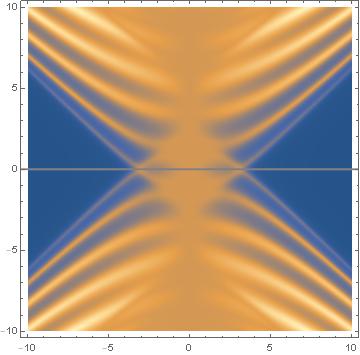}}
	\subfigure[$\omega=0$ $B_{ty}$ s]
	{\includegraphics[width=2.5cm]{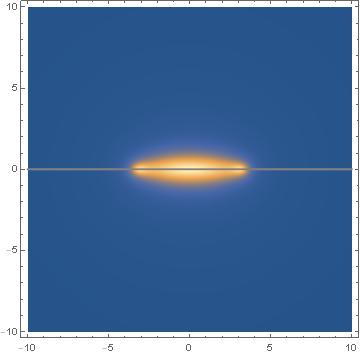}}
	\subfigure[$\omega=2$ $B_{ty}$ s]
	{\includegraphics[width=2.5cm]{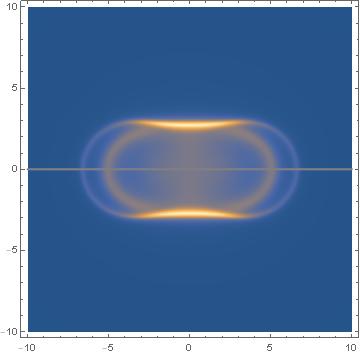}}
	\subfigure[c, $B_{ty(0)}$]
	{\includegraphics[width=2.5cm]{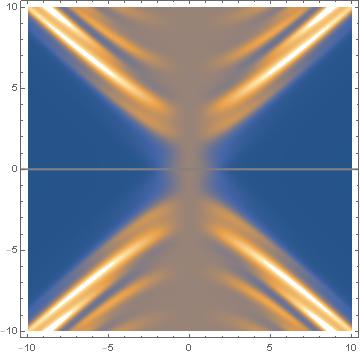}}
	\subfigure[$\omega=0$ $B_{ty}$ c]
	{\includegraphics[width=2.5cm]{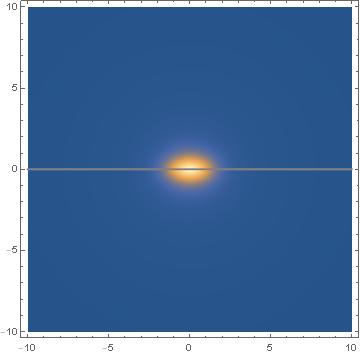}}
	\subfigure[$\omega=2$ $B_{ty}$ c]
	{\includegraphics[width=2.5cm]{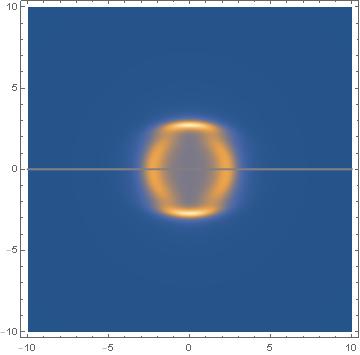}}\\
		\subfigure[$B_{rx(0)}$ s]
	{\includegraphics[width=2.5cm]{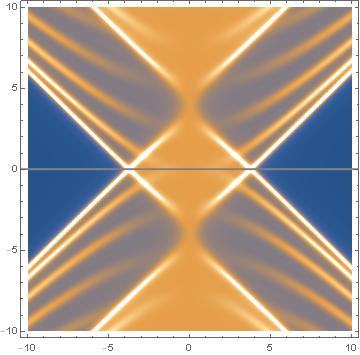}}
	\subfigure[$\omega=0$ $B_{rx}$ s]
	{\includegraphics[width=2.5cm]{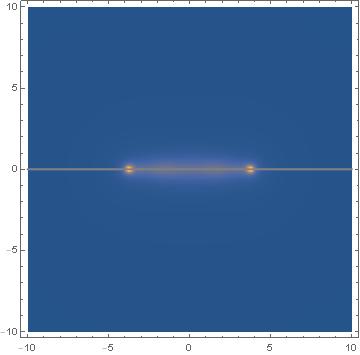}}
	\subfigure[$\omega=2$ $B_{rx}$ s]
	{\includegraphics[width=2.5cm]{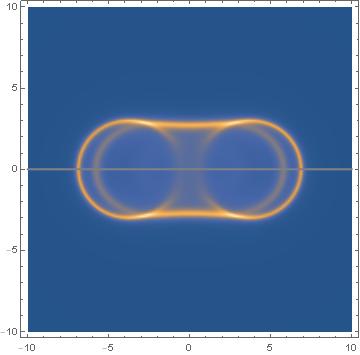}}
	\subfigure[c, $B_{rx(0)}$]
	{\includegraphics[width=2.5cm]{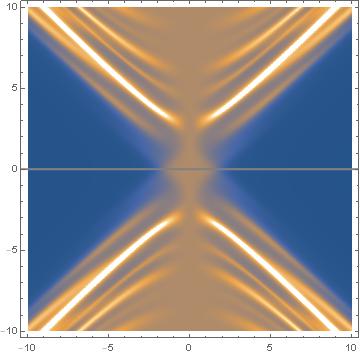}}
	\subfigure[$\omega=0$ $B_{rx}$ c]
	{\includegraphics[width=2.5cm]{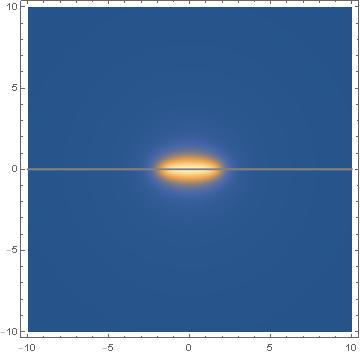}}
	\subfigure[$\omega=2$ $B_{rx}$ c]
	{\includegraphics[width=2.5cm]{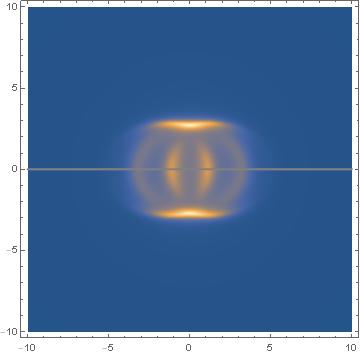}}
	\caption{Spectrum of  $B_{rx}$ and $B_{ty}$    with sliced views $k_{x}$-$k_{y}$ at $\omega=0,2$ slices, without which these spectra are ambiguous. Notice the zero modes and Ribbons connecting the two split cones. 
	}   \label{fig:Bx22_slice}
\end{figure}

  Notice the   manifest zero mode Disk in $B_{tr}$ from the figure \ref{fig:Bdd2_s}. There are rotational symmetry in $B_{tr},B_{xy}$, but notin $B_{xr}$. 
Spectrum of  $B_{ty}$ and $B_{xr}$ are ambiguous without 
 the   views in $k_{x}$-$k_{y}$ at various  $\omega$ slices, which we provide in figure \ref{fig:Bx22_slice}. Both of them have split cones and    the zero modes.  Both have Ribbon bands  connecting the two   cones. 
 
\section{Conclusion}
We   classified  the  Yukawa type interactions according to its Lorentz symmetry of the boundary theory and calculate their spectral functions.  
We   met many  interesting features that appear in the strongly correlated system:   gap, pseudogap, flat band and   Fermi-arc stuctures appear.   
   
 Out of the 16 interaction types, only parity breaking scalar  interaction  ($\Gamma=\mathbb{1}$) creates a gap without ambiguity. Both source and the condensation create gaps.    However,  the parity conserving  scalar  ($\Gamma=\Gamma^{5}$) has a zero mode Dirac cone in spectrum, which is much sharper than the case of the non-interacting case due to the transfer of the spectral weight  to the zero mode by  the interaction. The genuine  physical system with full gap will be described by this coupling. 
 For vector $B_{\mu}$,  $B_{5\mu}$,  $B_{r\mu}$ show feature of split cones if the order parameter field have source  part. 
The last two are invariant under the parity ,    there are the Fermi-arc. 
Another interesting feature is the flat disk band of $B_{xy}$, which might be useful to describe the twisted bi-layered graphene.  If  chemical  potential is applied, the disk bend like a bowl and the fermi level shifts, resembling the band of the Kondo lattice. 
 
 There are three classes of vectors,
$B_{\mu}, B_{r\mu}$ and $B_{5\mu}$, respectively. 
The source creates the splited cones and the condensation creates just asymmetry. 
The first two are invariant under the reflection showing  zero mode related features like  Fermi-arc and surface states(Ribbon band).
There are 3 tensor types: $\Gamma^{xy},\Gamma^{tx},\Gamma^{ty}$. These respect the parity and has zero modes.  

 Since the spectral data is the fingerprint of  a matter,   we would be able to determine the order of a strongly interacting system by comparing the calculated spectral function in the presence of  these orders   with the experimental data.  
We expect that our result will give an insight for  magnetic orders of strongly interacting materials. 

Final   remark is  that the quartic  or higher terms do not contribute to the spectral functions.Therefore it is enough to discuss the effect of the Yukawa coupling   to  calculate the leading order effects of order parameter fields on the fermion spectrum.  In fact,  the Yukawa coupling  terms are most relevant ones in low energy. Here we consider   the gravity background of  asymptotically AdS4  with Lorentz invariance. 

In the future, we will study the AdS$_{5}$ version of this paper, which   related to Weyl Semi-metal instead of Dirac semi-metal. 
It will be also interesting to extend our work to higher quantum critical points.  
There are ten classes of different topological insulator/superconductors depending on the discrete symmetries. It will be also interesting to realize all of such 10 folds way in terms of the explicit laglangian.  
One more possibility is to study the effect of  combinations of the Yukawa interactions to create different types of spectral features. 
Studies  in these directions are under progress.

 \appendix
      	\section{spectrum with one flavour}
	We present the results for 1 flavor, which in many case seems to give an half of the  2-flavor case with an asymmetry. 
\subsection{  Scalar }
\begin{figure}[ht!]
	\centering
	\subfigure[s $M_{0}$]
	{\includegraphics[width=3cm]{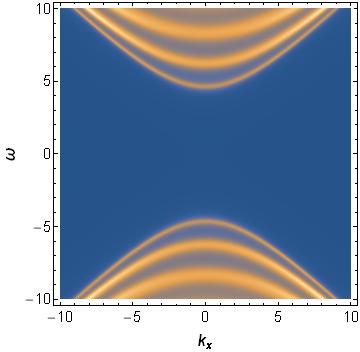}}
	\subfigure[s $M_{5}$]
	{\includegraphics[width=3cm]{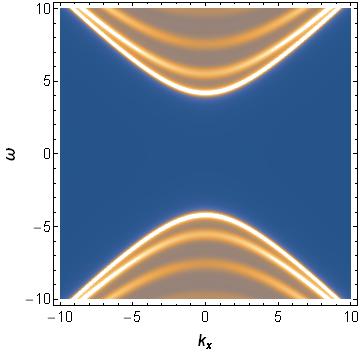}}
	\subfigure[c $M_{0}$]
	{\includegraphics[width=3cm]{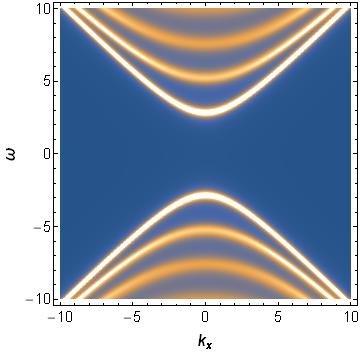}}
	\subfigure[c $M_{5}$]
	{\includegraphics[width=3cm]{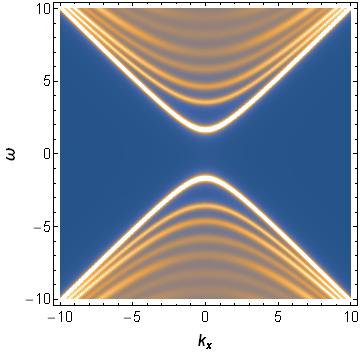}}
	\caption{SF (a) with scalar source,  Gap $\Delta\sim M_{0}$.   (b) with pseudo-scalar source. Gap  $\Delta\sim \sqrt{M}$.  (c) with scalar condensation  (d) with pseudo-scalar condensation. }
	\label{fig:scalar1_s} 
\end{figure}
Other type of scalars $\Gamma^{r5},\Gamma^{r}$ does not seem to change the spectrum of zero scalar coupling.  
Notice that there is no zero mode in one flavor scalar interaction. This is striking difference from the 2 flavor case. 

\subsection{Vectors}
\begin{figure}[ht!]  
	\centering	
 	\subfigure[s, $B_{x}$]
	{\includegraphics[width=2.5cm]{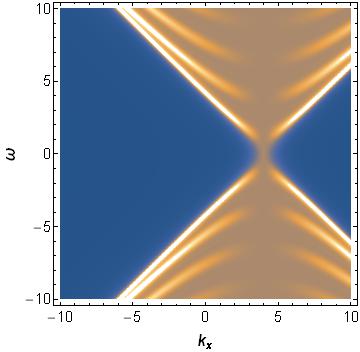}}
	\subfigure[$\omega=0$, $B_{x}$ s]
	{\includegraphics[width=2.5cm]{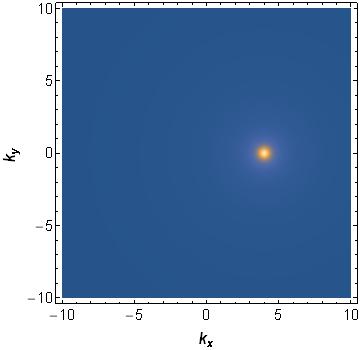}}
	\subfigure[$\omega=2$, $B_{x}$ s]
	{\includegraphics[width=2.5cm]{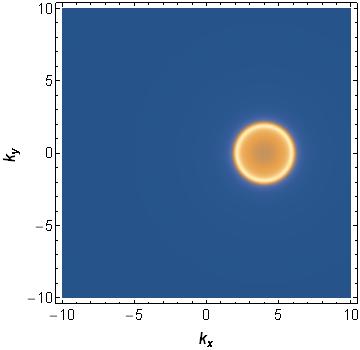}}
	\subfigure[c, $B_{x}$]
	{\includegraphics[width=2.5cm]{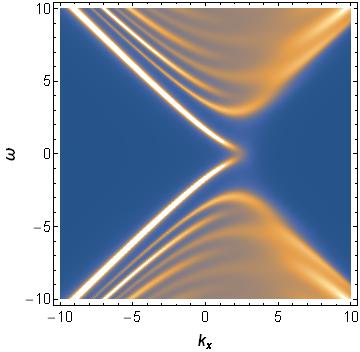}}
	\subfigure[$\omega=0$, $B_{x}$ c]
	{\includegraphics[width=2.5cm]{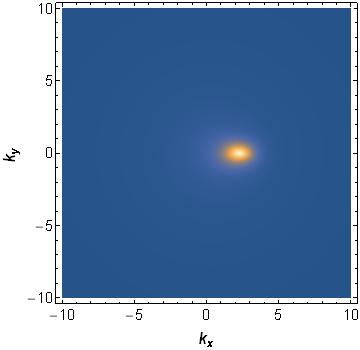}}
	\subfigure[$\omega=2$, $B_{x}$ c]
	{\includegraphics[width=2.5cm]{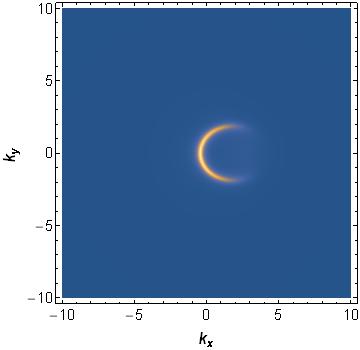}}\\
	\subfigure[s, $B_{5x}$]
	{\includegraphics[width=2.5cm]{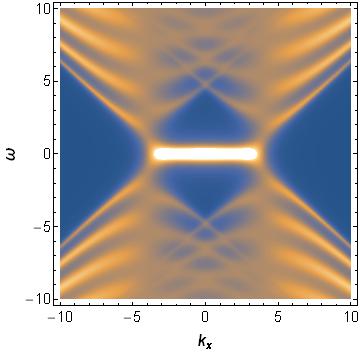}}
	\subfigure[$\omega=0$, $B_{5x}$ s]
	{\includegraphics[width=2.5cm]{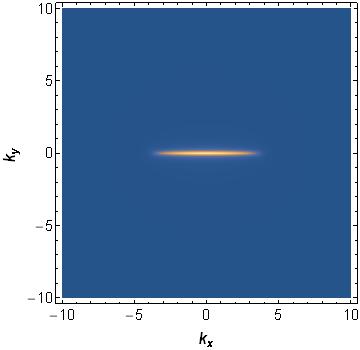}}
	\subfigure[$\omega=2$, $B_{5x}$ s]
	{\includegraphics[width=2.5cm]{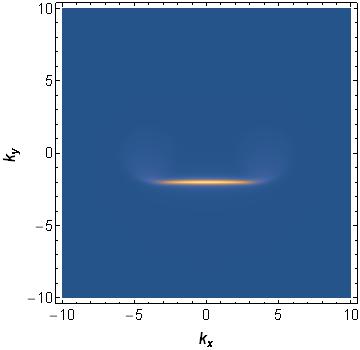}}
	\subfigure[c, $B_{5x}$]
	{\includegraphics[width=2.5cm]{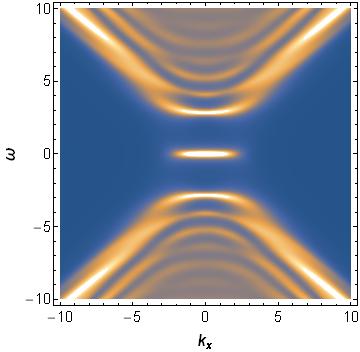}}
	\subfigure[$\omega=0$, $B_{5x}$ c]
	{\includegraphics[width=2.5cm]{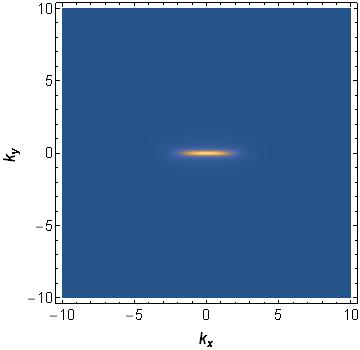}}
	\subfigure[$\omega=2$, $B_{5x}$ c]
	{\includegraphics[width=2.5cm]{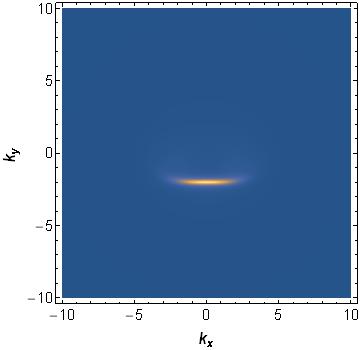}}
	\caption{SF with most typical     vector   interactions}   \label{fig:Bx1_s}
\end{figure}

\begin{figure}[ht!]
	\centering	
	\subfigure[s, $B_{t(0)}$]
	{\includegraphics[width=2.5cm]{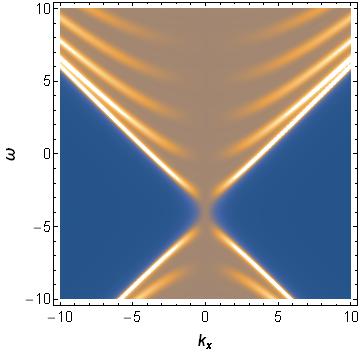}}
	\subfigure[s, $B_{y(0)}$]
	{\includegraphics[width=2.5cm]{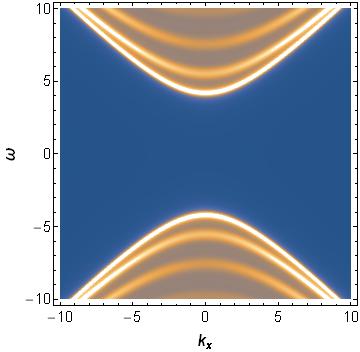}}
	\subfigure[s, $B_{5t(0)}$]
	{\includegraphics[width=2.5cm]{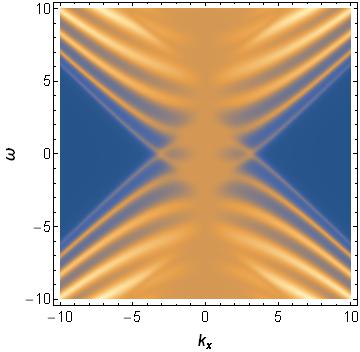}}
	\subfigure[s, $B_{5y(0)}$]
	{\includegraphics[width=2.5cm]{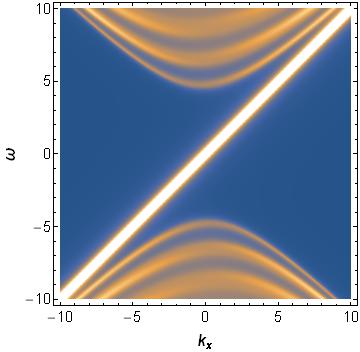}}
	\\
	\subfigure[c, $B_{t(0)}$]
	{\includegraphics[width=2.5cm]{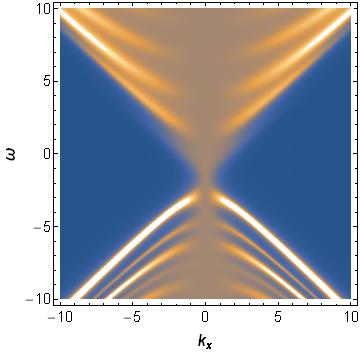}}
	\subfigure[c, $B_{y(0)}$]
	{\includegraphics[width=2.5cm]{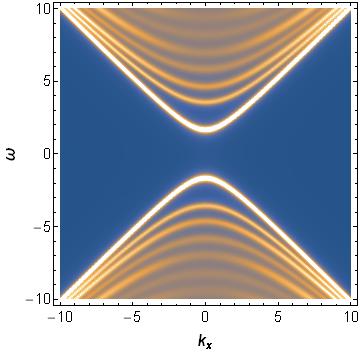}}
	\subfigure[c, $B_{5t(0)}$]
	{\includegraphics[width=2.5cm]{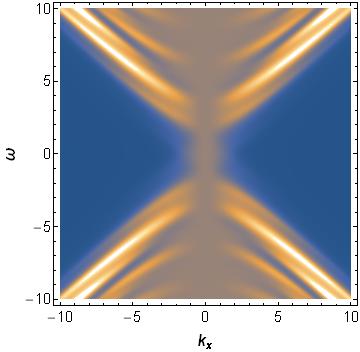}}
	\subfigure[c, $B_{5y(0)}$]
	{\includegraphics[width=2.5cm]{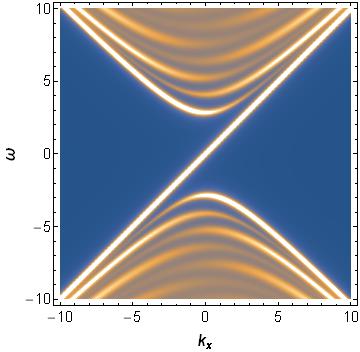}}
	\caption{SF with various types vector condensation interaction}   \label{fig:Bx1_c}
\end{figure}

%
%
%

\newpage
\subsection{Anti-symmetric Tensor }
 
\begin{figure}[ht!]
	\centering	
	\centering	
	\subfigure[s, $B_{ty(0)}$]
	{\includegraphics[width=2.5cm]{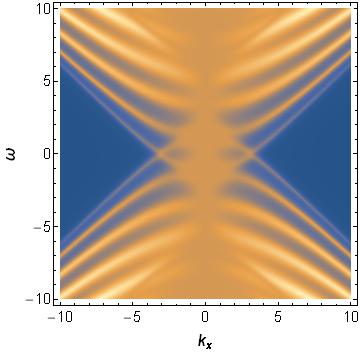}}
	\subfigure[$\omega=0$ $B_{ty}$ s]
	{\includegraphics[width=2.5cm]{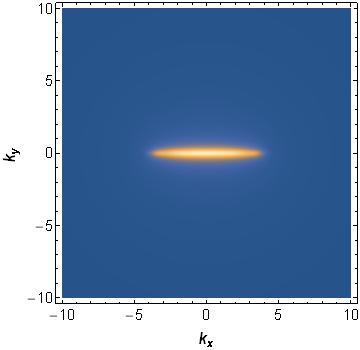}}
	\subfigure[$\omega=2$ $B_{ty}$ s]
	{\includegraphics[width=2.5cm]{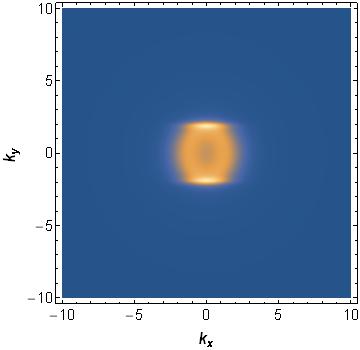}}
	\subfigure[c, $B_{ty(0)}$]
	{\includegraphics[width=2.5cm]{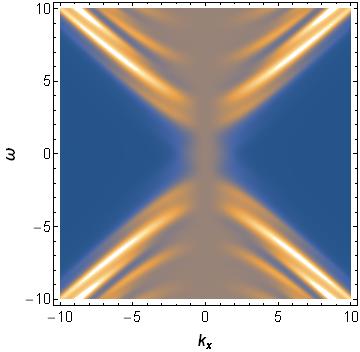}}
	\subfigure[$\omega=0$ $B_{ty}$ c]
	{\includegraphics[width=2.5cm]{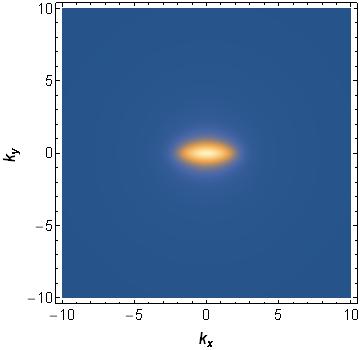}}
	\subfigure[$\omega=2$ $B_{ty}$ c]
	{\includegraphics[width=2.5cm]{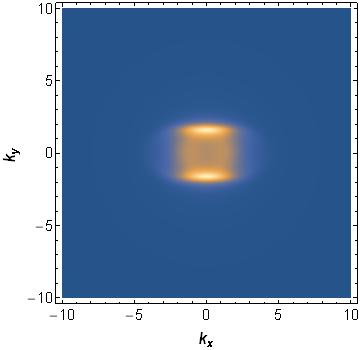}}
	\subfigure[s, $B_{xr(0)}$]
	{\includegraphics[width=2.5cm]{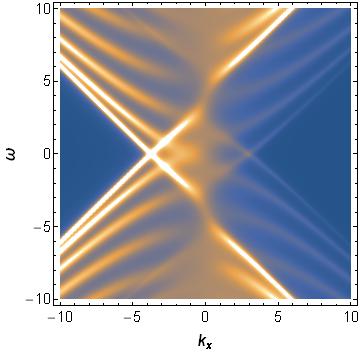}}
	\subfigure[$\omega=0$ $B_{xr}$ s]
	{\includegraphics[width=2.5cm]{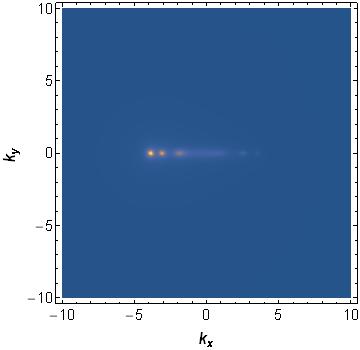}}
	\subfigure[$\omega=2$ $B_{xr}$ s]
	{\includegraphics[width=2.5cm]{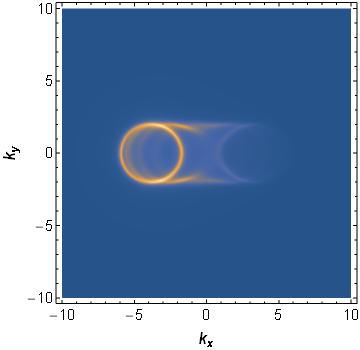}}
	\subfigure[c, $B_{xr(0)}$]
	{\includegraphics[width=2.5cm]{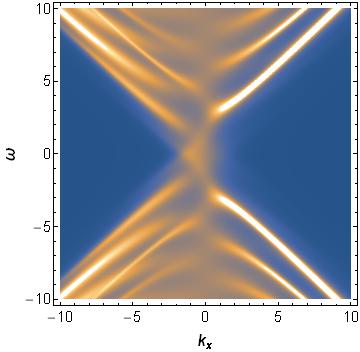}}
	\subfigure[$\omega=0$ $B_{xr}$ c]
	{\includegraphics[width=2.5cm]{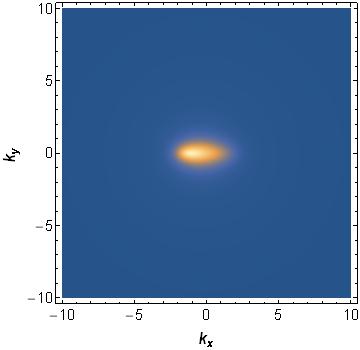}}
	\subfigure[$\omega=2$ $B_{xr}$ c]
	{\includegraphics[width=2.5cm]{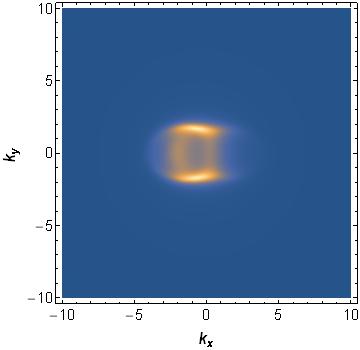}}
	\\
		\subfigure[s, $B_{xy}$]
	{\includegraphics[width=2.5cm]{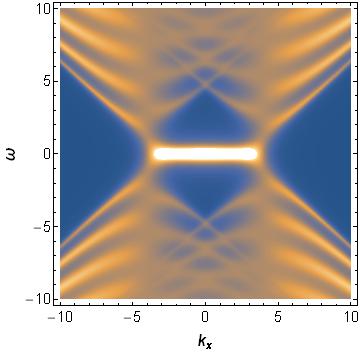}}
	\subfigure[s, $B_{tr }$]
	{\includegraphics[width=2.5cm]{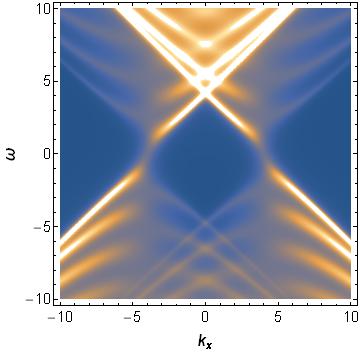}}
	\subfigure[c, $B_{xy}$]
	{\includegraphics[width=2.5cm]{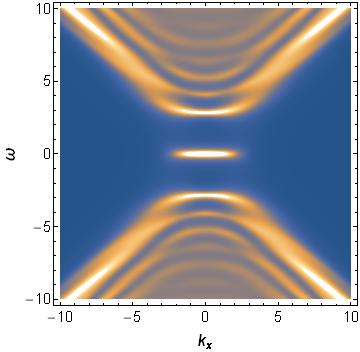}}
	\subfigure[c, $B_{tr}$]
	{\includegraphics[width=2.5cm]{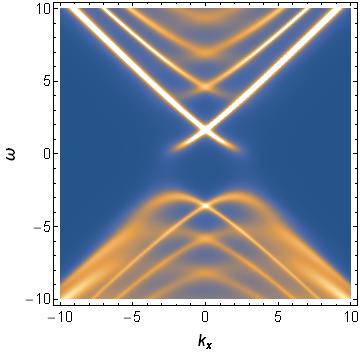}}
	\caption{Spectrum of  Anti-symmetric tensors.
	}   \label{fig:Bx1_slice}
\end{figure}

\section{The role of the chemical potential}   
\begin{figure}[ht!]
		\centering
		\subfigure[$\mu=0$, $A(w,k)$]
		{\includegraphics[width=3cm]{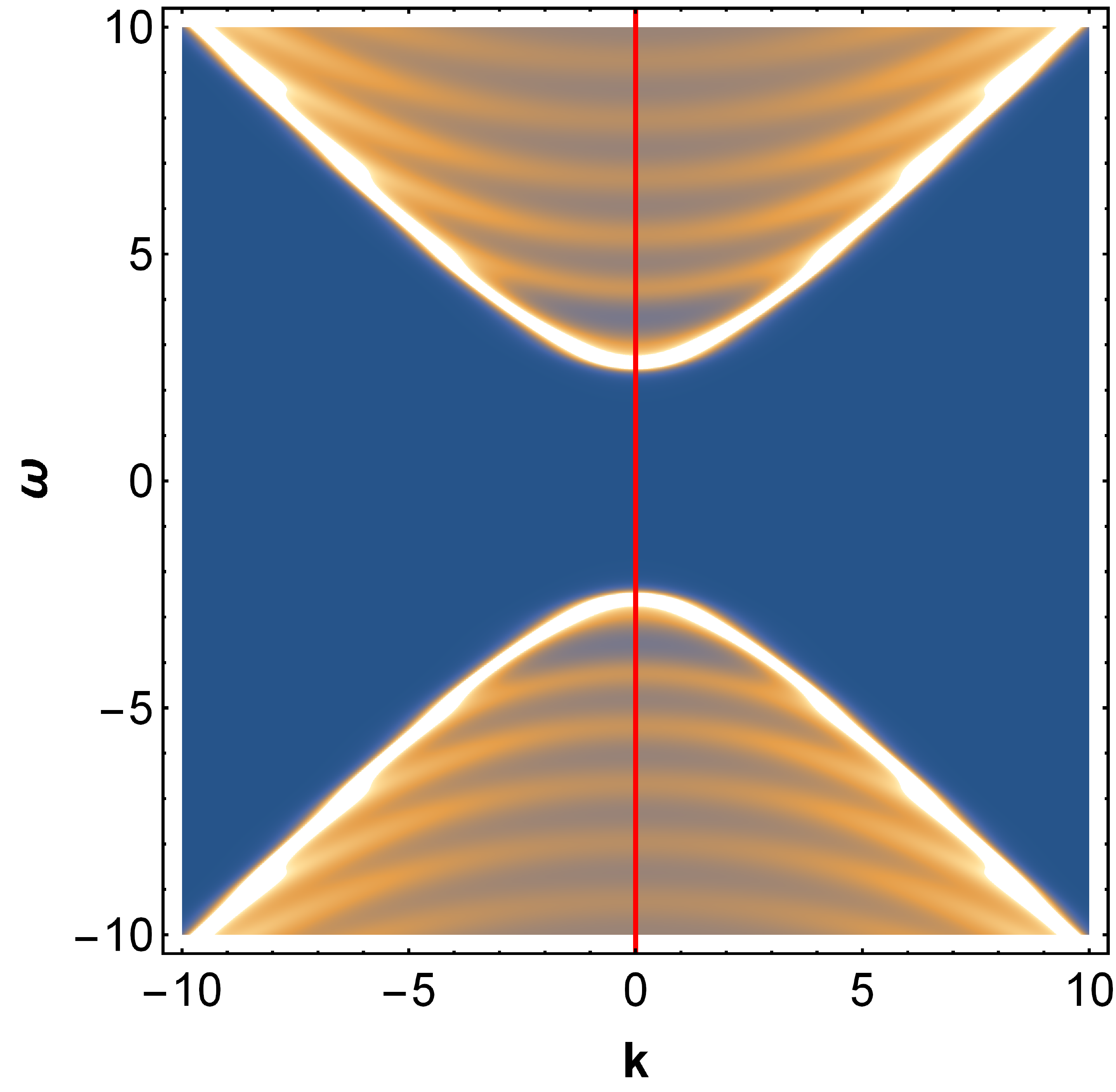}}
		\subfigure[$\mu=2\sqrt{3}$, $A(w,k)$]
		{\includegraphics[width=3cm]{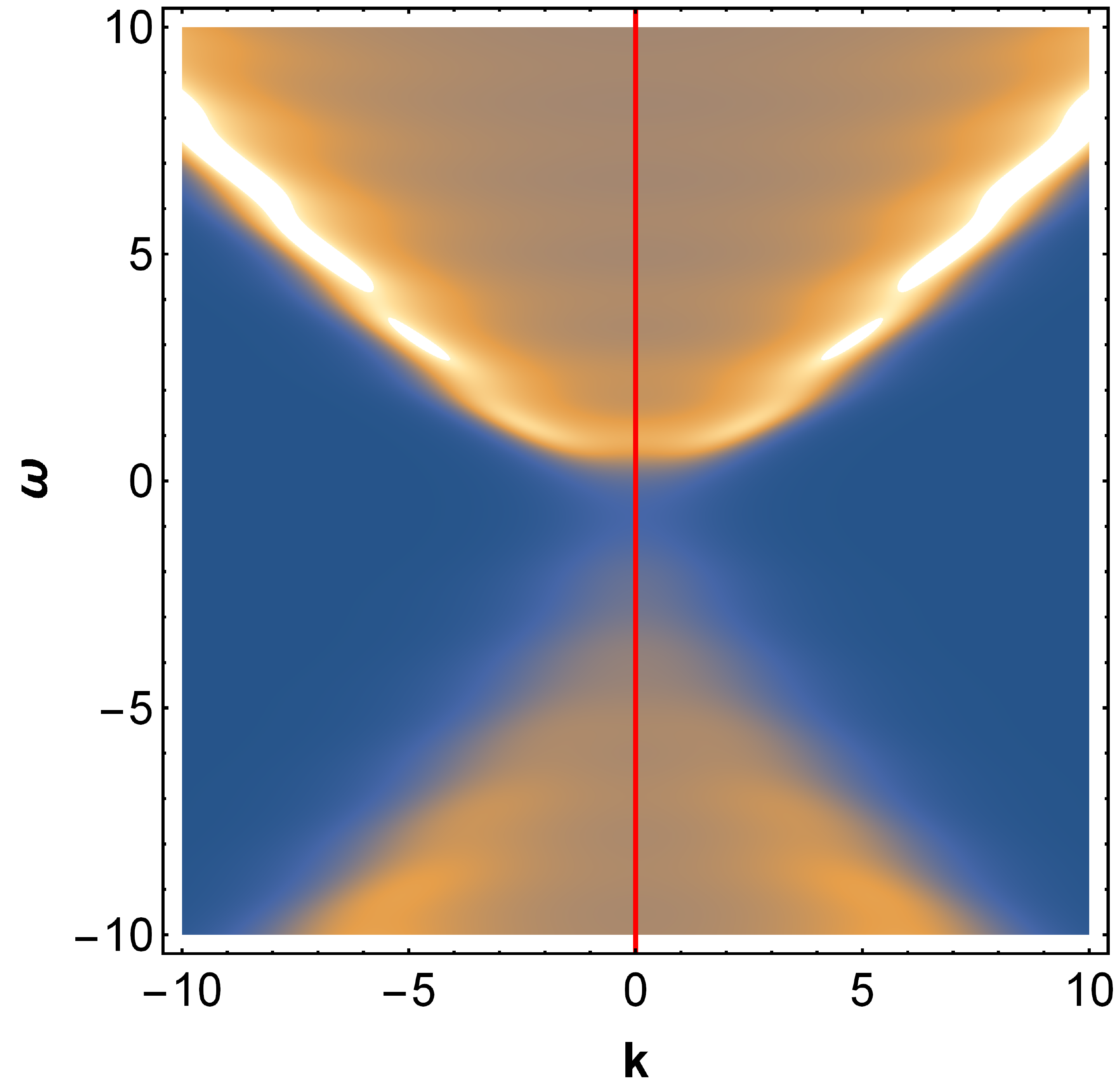}}
		\subfigure[$A(w,k=0)$]
		{\includegraphics[width=4.5cm]{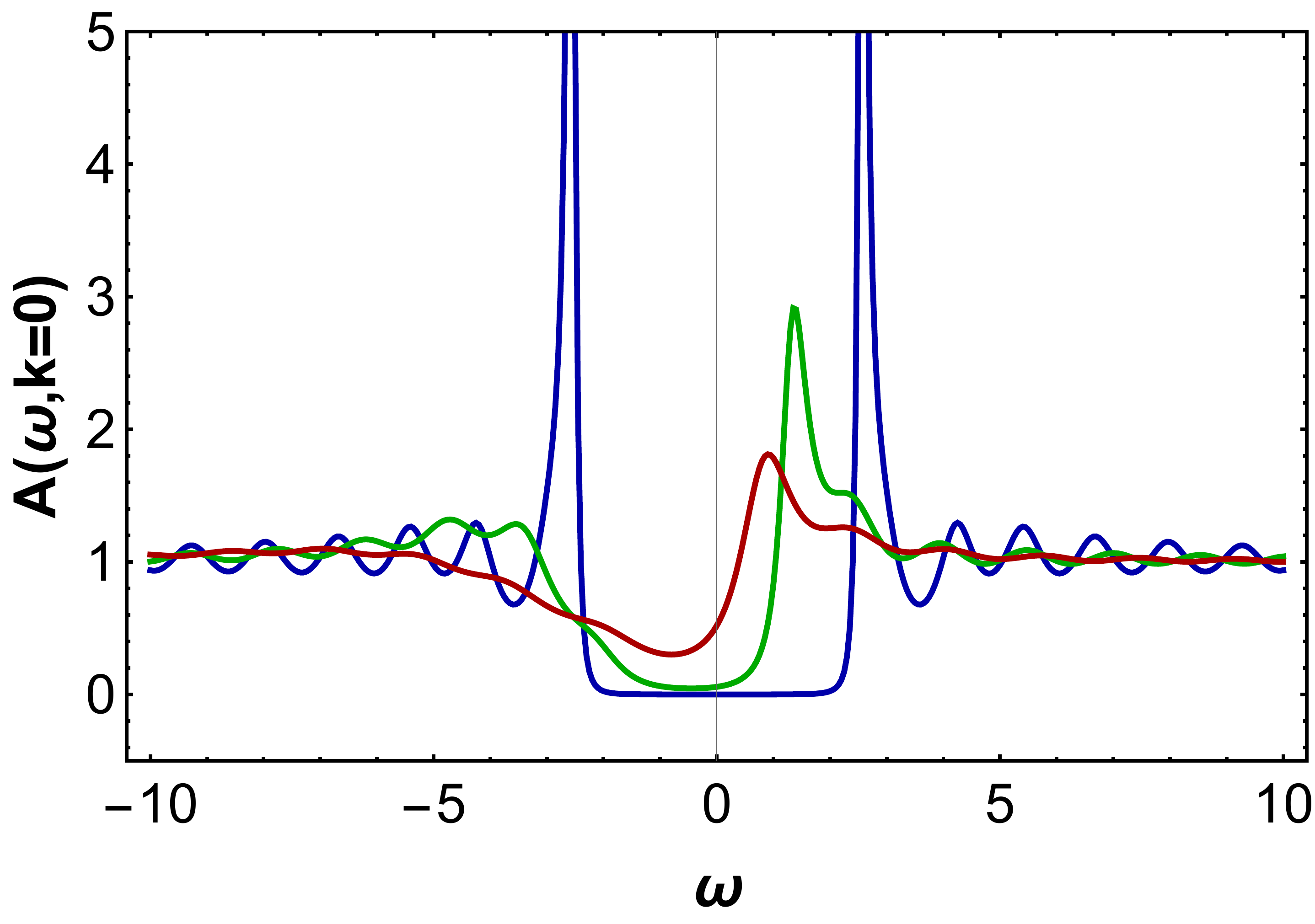}}
		\caption{ The role of $\mu$ for the case with  scalar condensation :    (a)   without and (b) with  chemical potential.  (c) Spectrum at  $k=0$:  blue for $\mu=0$, green for $\mu=\sqrt{3}$, red  for $\mu=2 \sqrt{3}$. 
 }\label{fig:chemical}
	\end{figure}

Let's begin by looking at the simplest  scalar case. As one can see from figure \ref{fig:chemical}, the main effect of the chemical potential is three folds: the first is to shift the Fermi-level and the second is to make the spectrum fuzzier. The third one is to introduce the asymmetry between the positive and negative frequency regions. 

\begin{figure}[ht!]
\centering
\subfigure[$B_{xy}$ s, $\mu=0$]
{\includegraphics[width=3cm]{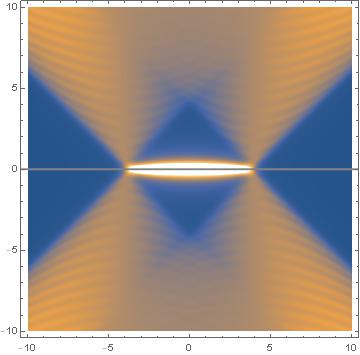}}
\subfigure[$B_{xy}$ s, $\mu=\sqrt{3}$]
{\includegraphics[width=3cm]{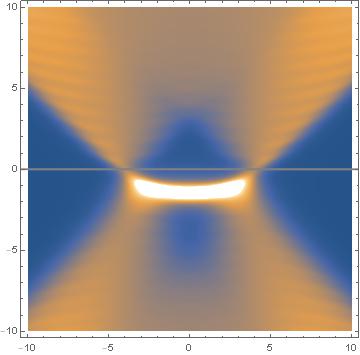}}
\subfigure[$B_{xy}$ c, $\mu=0$]
{\includegraphics[width=3cm]{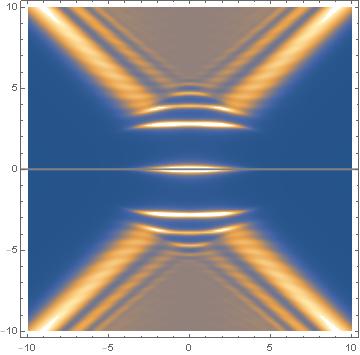}}
\subfigure[$B_{xy}$ c, $\mu=\sqrt{3}$]
{\includegraphics[width=3cm]{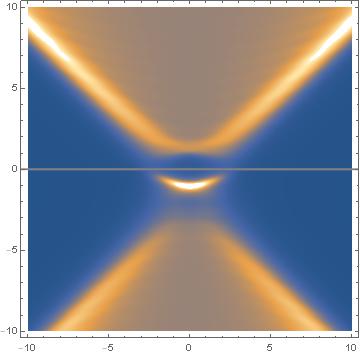}}
\caption{(a,b): SF with  only source $B_{xy}$ with one flavor. Notice   the presence of the  flat band of disk-shape, which is bent down in the presence of chemical potential; 
(c,d): SF with only condensation $B_{xy}$. 
 All figures here have cylindrical symmetry. }   
		\label{fig:Bxy}
	\end{figure}

 Now we consider the one flavor $B_{xy}$ and $B_{tr}$  interactions which are in a sense dual to each other. 
  To consider the case of spontaneously broken symmetry, we set $B_{(-1)}=0$. Most important feature here is the appearance of the flat band and gated gap, which means that the gap is reachable by gating. 
For $\mu=0$, exact flat band is generated at $w=0$. If we turn on the chemical potential, the central flat band is bent to give a shallow-bowl shaped  band. With increasing  A fuzzy flat band is created at higher energy. See Figure \ref{fig:Bxy}(d). 
Notice the  similarity  of the spectrum with that of the heavy fermion spectrum in  Kondo lattice, 
 where an impurity   spin  interact with the itinerant electron   with anti-ferromagnetic  coupling. 
Indeed, our interaction term $\boldmath{B_{xy}}{\bar \psi}\Gamma^{xy}\psi$  is   the form of the Kondo coupling $\boldmath{\vec{S}_{imp}}\cdot{\bar \psi}\vec{\sigma}\psi$  in the boundary  if we interpret the order parameter field $B_{xy}$ as the impurity spin along $z$ axis, because   $\Gamma^{xy}$ is the corresponding spin  generator matrix.  	
	%


 %
Both source and condensation seem to   develop  a  gap    in the negative energy region while Fermi level passes through the conduction band. Since  gap usually means gap containing the Fermi-level to give an insulator, we need a new name.  We call such gap as gated gap because we expect that we can get an insulator by gating the system, that is by applying external electric field.  
 Therefore we studied the  { evolution of $B_{rt}$} in $\mu$. 
 Figure \ref{fig:Brt_mu} shows, however,  that gap is   generated  only in a window of negative $\mu$ around $\mu=-1$.

\begin{figure}[ht!]
	\centering
	\subfigure[$\mu=-2\sqrt{3}$]
	{\includegraphics[width=3cm]{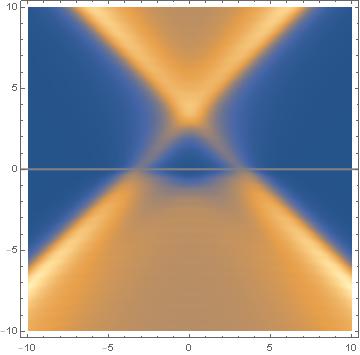}}
	\subfigure[$\mu=-\sqrt{3}$]
	{\includegraphics[width=3cm]{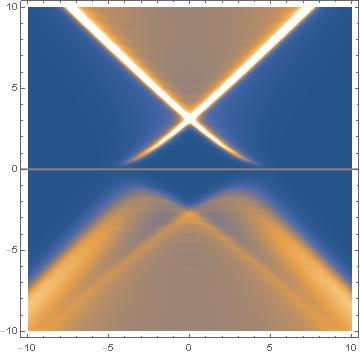}}
	\subfigure[$\mu=0$]
	{\includegraphics[width=3cm]{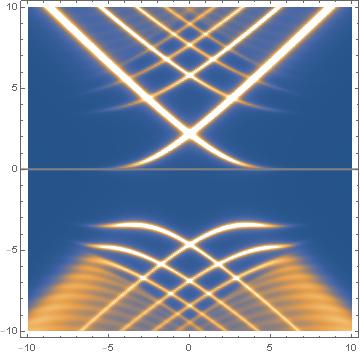}}
	\subfigure[$\mu=\sqrt{3}$]
	{\includegraphics[width=3cm]{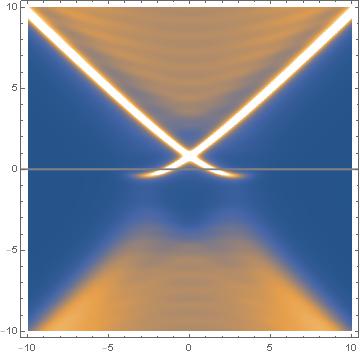}}
	\subfigure[$\mu=2\sqrt{3}$]
	{\includegraphics[width=3cm]{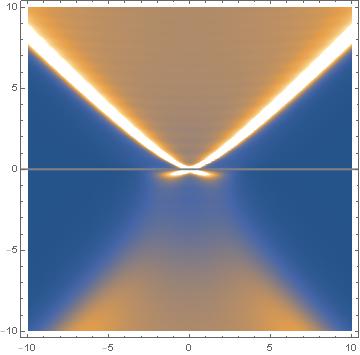}}
	\subfigure[$\mu=-2\sqrt{3}$ to $\mu=2\sqrt{3}$]
	{\includegraphics[width=10cm]{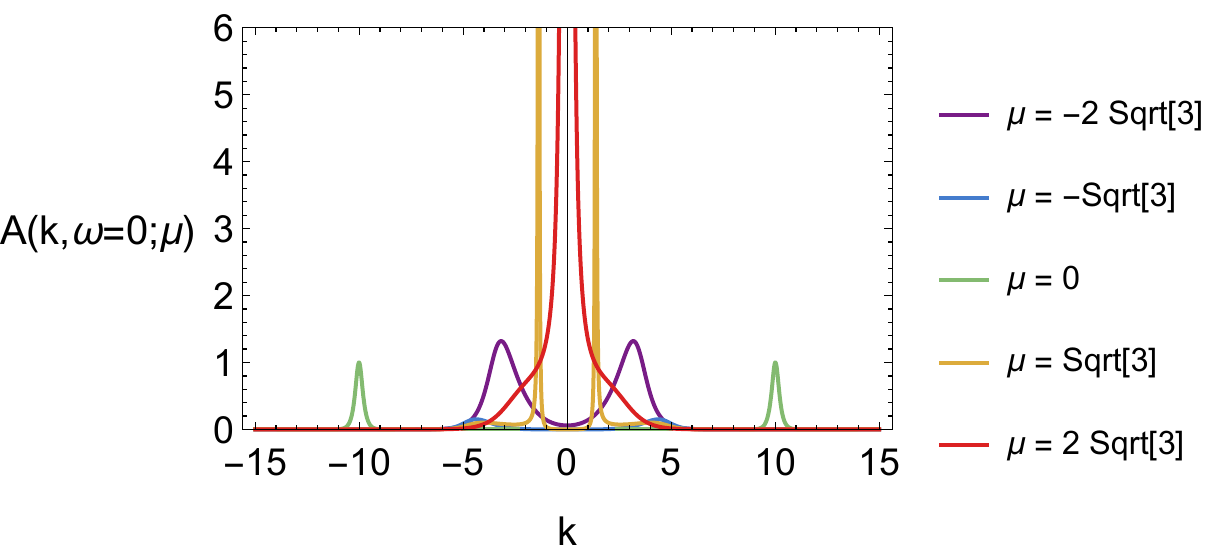}}
	\caption{$\mu$-evolution of SF of $B_{rt}$ interaction with condensation $B_{rt(0)}=7$.
	 (a,b,c,d,e): denisity plot in $(\omega,k)$. (f):SF at the Fermi level.  	The minimum of DOS is at $\mu \simeq -1$. } \label{fig:Brt_mu}
\end{figure}
 
	 \subsection{$B_{rt}$ vs $F_{rt}$:} We want to make a comment on  comparing this case with the work of Phillips et.al\cite{Edalati:2010ww,Edalati:2010ge} where the dipole interaction $F_{rt}{\bar \psi} \Gamma^{rt}\psi$   was considered in search of the gap. 
	The difference with ours is that our $B_{rt}$ is an order parameter  independent of the basic vector field $A_{\mu}$ while $F_{rt}=\partial_{r}A_{t}$. Namely, 
\begin{align}
A&=\mu_1- \frac{r_0}{r})dt, \quad 
F=dA=\frac {\mu r_{0}}{r^2}dt\wedge dr, \quad 
B= B^{(0)}_{rt}  dt\wedge dr. 
\end{align}
If the condensation $B^{(0)}_{rt}$ is related to the chemical potential by $B^{(0)}_{rt}=\mu/r_{0}$,  our model with  non-zero chemical potential is    exactly the same as the dipole interaction model. In fact, Figure \ref{fig:Brt_mu} shows that the overall features of two cases are similar.  
However, for zero chemical potential,  our model  still have the non-zero order parameter  $B^{(0)}_{rt}$ while  the dipole interaction  vanishes automatically. \begin{figure}[ht!]
	\centering
	\subfigure[$ \mu=\sqrt{3},F_{rt}$]
	{\includegraphics[width=3cm]{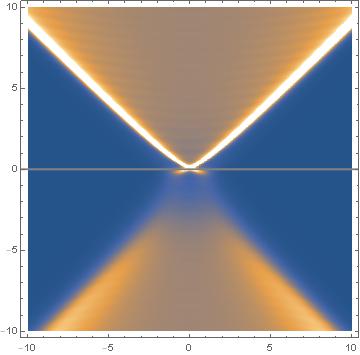}}
	\subfigure[$\mu=2\sqrt{3}, F_{rt}$]
	{\includegraphics[width=3cm]{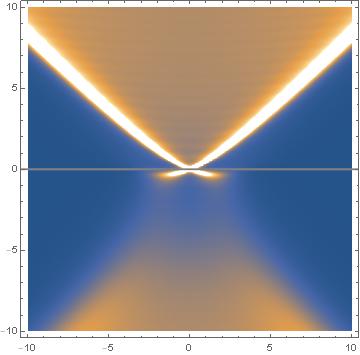}}
	\subfigure[$\mu=\sqrt{3},B_{rt}$]
	{\includegraphics[width=3cm]{./fig/brt/m_0_Btr_7_.jpg}}
	\subfigure[$\mu=2\sqrt{3}, B_{rt}$]
	{\includegraphics[width=3cm]{./fig/brt/mu_2_Btr_7_.jpg}}
	\caption{ (a,b): SF for $F_{rt}$ and (c,d): SF for $B_{rt}$ with condensation  $B^{(0)}_{rt}=7$. Two models are similar.  
	}   \label{fig:Brt_mu}
\end{figure}
Both source and condensation seem to   develop  a  gap    in the negative energy region while Fermi level passes through the conduction band. Since  gap usually means gap containing the Fermi-level to give an insulator, we need a new name.  We call such gap as gated gap because we expect that we can get an insulator by gating the system, that is by applying external electric field.  
 Therefore we studied the  { evolution of $B_{rt}$} in $\mu$. 
 Figure \ref{fig:Brt_mu} shows, however,  that gap is   generated  only in a window of negative $\mu$ around $\mu=-1$.

\newpage

  \acknowledgments
 This  work is supported by Mid-career Researcher Program through the National Research Foundation of Korea grant No. NRF-2016R1A2B3007
687.   YS was   supported by Basic Science Research Program through NRF grant No. NRF-2019R1I1A1A01057998.   We also would like to
thank the APCTP  focus program, “Quantum Matter from the Entanglement and Holography” in Pohang, Korea for the hospitality
during our visit, where part of this work was done.
 

\bibliographystyle{JHEP}
\bibliography{Refs_SO.bib}

\providecommand{\href}[2]{#2}\begingroup\raggedright\begin{thebibliography}{10}

\bibitem{pkim}
J.~{Crossno}, J.~K. {Shi}, K.~{Wang}, X.~{Liu}, A.~{Harzheim}, A.~{Lucas}
  et~al., \emph{{Observation of the Dirac fluid and the breakdown of the
  Wiedemann-Franz law in graphene}},
  \href{http://dx.doi.org/10.1126/science.aad0343}{\emph{Science} {\bf 351}
  (Mar., 2016) 1058--1061}, [\href{http://arxiv.org/abs/1509.04713}{{\tt
  1509.04713}}].

\bibitem{Lucas:2015sya}
A.~Lucas, J.~Crossno, K.~C. Fong, P.~Kim and S.~Sachdev, \emph{{Transport in
  inhomogeneous quantum critical fluids and in the Dirac fluid in graphene}},
  \href{http://dx.doi.org/10.1103/PhysRevB.93.075426}{\emph{Phys. Rev.} {\bf
  B93} (2016) 075426}, [\href{http://arxiv.org/abs/1510.01738}{{\tt
  1510.01738}}].

\bibitem{liu2012crossover}
M.~Liu, J.~Zhang, C.-Z. Chang, Z.~Zhang, X.~Feng, K.~Li et~al., \emph{Crossover
  between weak antilocalization and weak localization in a magnetically doped
  topological insulator}, {\emph{Phys. Rev. Lett.} {\bf 108} (2012) 036805}.

\bibitem{zhang2012interplay}
D.~Zhang et~al., \emph{Interplay between ferromagnetism, surface states, and
  quantum corrections in a magnetically doped topological insulator},
  {\emph{Physical Review B} {\bf 86} (2012) 205127}.

\bibitem{bao2013quantum}
L.~Bao, W.~Wang, N.~Meyer, Y.~Liu, C.~Zhang, K.~Wang et~al., \emph{Quantum
  corrections crossover and ferromagnetism in magnetic topological insulators},
  {\emph{Scientific reports} {\bf 3} (2013) }.

\bibitem{Seo:2016vks}
Y.~Seo, G.~Song, P.~Kim, S.~Sachdev and S.-J. Sin, \emph{{Holography of the
  Dirac Fluid in Graphene with two currents}},
  \href{http://dx.doi.org/10.1103/PhysRevLett.118.036601}{\emph{Phys. Rev.
  Lett.} {\bf 118} (2017) 036601}, [\href{http://arxiv.org/abs/1609.03582}{{\tt
  1609.03582}}].

\bibitem{Seo:2017oyh}
Y.~Seo, G.~Song and S.-J. Sin, \emph{{Strong Correlation Effects on Surfaces of
  Topological Insulators via Holography}},
  \href{http://dx.doi.org/10.1103/PhysRevB.96.041104}{\emph{Phys. Rev.} {\bf
  B96} (2017) 041104}, [\href{http://arxiv.org/abs/1703.07361}{{\tt
  1703.07361}}].

\bibitem{Seo:2017yux}
Y.~Seo, G.~Song, C.~Park and S.-J. Sin, \emph{{Small Fermi Surfaces and Strong
  Correlation Effects in Dirac Materials with Holography}},
  \href{http://dx.doi.org/10.1007/JHEP10(2017)204}{\emph{JHEP} {\bf 10} (2017)
  204}, [\href{http://arxiv.org/abs/1708.02257}{{\tt 1708.02257}}].

\bibitem{Coleman:2015uma}
P.~Coleman, \emph{{Heavy Fermions and the Kondo Lattice: a 21st Century
  Perspective}},  2015.
\newblock \href{http://arxiv.org/abs/1509.05769}{{\tt 1509.05769}}.

\bibitem{cao2018unconventional}
Y.~Cao, V.~Fatemi, S.~Fang, K.~Watanabe, T.~Taniguchi, E.~Kaxiras et~al.,
  \emph{Unconventional superconductivity in magic-angle graphene
  superlattices}, {\emph{Nature} {\bf 556} (2018) 43--50}.

\bibitem{cao2018correlated}
Y.~Cao, V.~Fatemi, A.~Demir, S.~Fang, S.~L. Tomarken, J.~Y. Luo et~al.,
  \emph{Correlated insulator behaviour at half-filling in magic-angle graphene
  superlattices}, {\emph{Nature} {\bf 556} (2018) 80}.

\bibitem{wilson1971renormalization}
K.~G. Wilson, \emph{Renormalization group and critical phenomena. i.
  renormalization group and the kadanoff scaling picture}, {\emph{Physical
  review B} {\bf 4} (1971) 3174}.

\bibitem{wilson1975renormalization}
K.~G. Wilson, \emph{The renormalization group: Critical phenomena and the kondo
  problem}, {\emph{Reviews of modern physics} {\bf 47} (1975) 773}.

\bibitem{Maldacena:1997re}
J.~M. Maldacena, \emph{{The Large N limit of superconformal field theories and
  supergravity}},
  \href{http://dx.doi.org/10.1023/A:1026654312961}{\emph{Int.J.Theor.Phys.}
  {\bf 38} (1999) 1113--1133}, [\href{http://arxiv.org/abs/hep-th/9711200}{{\tt
  hep-th/9711200}}].

\bibitem{Witten:1998qj}
E.~Witten, \emph{{Anti-de Sitter space and holography}}, {\emph{Adv. Theor.
  Math. Phys.} {\bf 2} (1998) 253--291},
  [\href{http://arxiv.org/abs/hep-th/9802150}{{\tt hep-th/9802150}}].

\bibitem{Gubser:1998bc}
S.~S. Gubser, I.~R. Klebanov and A.~M. Polyakov, \emph{{Gauge theory
  correlators from noncritical string theory}},
  \href{http://dx.doi.org/10.1016/S0370-2693(98)00377-3}{\emph{Phys. Lett.}
  {\bf B428} (1998) 105--114}, [\href{http://arxiv.org/abs/hep-th/9802109}{{\tt
  hep-th/9802109}}].

\bibitem{alvarez1998geometric}
E.~Alvarez and C.~Gomez, \emph{Geometric holography, the renormalization group
  and the c-theorem}, {\emph{arXiv preprint hep-th/9807226} (1998) }.

\bibitem{balasubramanian1999spacetime}
V.~Balasubramanian and P.~Kraus, \emph{Spacetime and the holographic
  renormalization group}, {\emph{Physical Review Letters} {\bf 83} (1999)
  3605}.

\bibitem{de2000holographic}
J.~De~Boer, E.~Verlinde and H.~Verlinde, \emph{On the holographic
  renormalization group}, {\emph{Journal of High Energy Physics} {\bf 2000}
  (2000) 003}.

\bibitem{heemskerk2011holographic}
I.~Heemskerk and J.~Polchinski, \emph{Holographic and wilsonian renormalization
  groups}, {\emph{Journal of High Energy Physics} {\bf 2011} (2011) 31}.

\bibitem{fradkin2013field}
E.~Fradkin, \emph{Field theories of condensed matter physics}.
\newblock Cambridge University Press, 2013.

\bibitem{sslee}
S.-S. Lee, \emph{{A Non-Fermi Liquid from a Charged Black Hole: A Critical
  Fermi Ball}}, \href{http://dx.doi.org/10.1103/PhysRevD.79.086006}{\emph{Phys.
  Rev.} {\bf D79} (2009) 086006}, [\href{http://arxiv.org/abs/0809.3402}{{\tt
  0809.3402}}].

\bibitem{Liu:2009dm}
H.~Liu, J.~McGreevy and D.~Vegh, \emph{{Non-Fermi liquids from holography}},
  \href{http://dx.doi.org/10.1103/PhysRevD.83.065029}{\emph{Phys. Rev.} {\bf
  D83} (2011) 065029}, [\href{http://arxiv.org/abs/0903.2477}{{\tt
  0903.2477}}].

\bibitem{Iqbal:2009fd}
N.~Iqbal and H.~Liu, \emph{{Real-time response in AdS/CFT with application to
  spinors}}, \href{http://dx.doi.org/10.1002/prop.200900057}{\emph{Fortsch.
  Phys.} {\bf 57} (2009) 367--384}, [\href{http://arxiv.org/abs/0903.2596}{{\tt
  0903.2596}}].

\bibitem{Cubrovic:2009ye}
M.~Cubrovic, J.~Zaanen and K.~Schalm, \emph{{String Theory, Quantum Phase
  Transitions and the Emergent Fermi-Liquid}},
  \href{http://dx.doi.org/10.1126/science.1174962}{\emph{Science} {\bf 325}
  (2009) 439--444}, [\href{http://arxiv.org/abs/0904.1993}{{\tt 0904.1993}}].

\bibitem{Faulkner:2009am}
T.~Faulkner, G.~T. Horowitz, J.~McGreevy, M.~M. Roberts and D.~Vegh,
  \emph{{Photoemission 'experiments' on holographic superconductors}},
  \href{http://dx.doi.org/10.1007/JHEP03(2010)121}{\emph{JHEP} {\bf 03} (2010)
  121}, [\href{http://arxiv.org/abs/0911.3402}{{\tt 0911.3402}}].

\bibitem{benini2011holographic}
F.~Benini, C.~P. Herzog and A.~Yarom, \emph{Holographic fermi arcs and a d-wave
  gap}, {\emph{Physics Letters B} {\bf 701} (2011) 626--629}.

\bibitem{Vegh:2010fc}
D.~Vegh, \emph{{Fermi arcs from holography}},
  \href{http://arxiv.org/abs/1007.0246}{{\tt 1007.0246}}.

\bibitem{Edalati:2010ww}
M.~Edalati, R.~G. Leigh and P.~W. Phillips, \emph{{Dynamically Generated Mott
  Gap from Holography}},
  \href{http://dx.doi.org/10.1103/PhysRevLett.106.091602}{\emph{Phys. Rev.
  Lett.} {\bf 106} (2011) 091602}, [\href{http://arxiv.org/abs/1010.3238}{{\tt
  1010.3238}}].

\bibitem{Edalati:2010ge}
M.~Edalati, R.~G. Leigh, K.~W. Lo and P.~W. Phillips, \emph{{Dynamical Gap and
  Cuprate-like Physics from Holography}},
  \href{http://dx.doi.org/10.1103/PhysRevD.83.046012}{\emph{Phys. Rev.} {\bf
  D83} (2011) 046012}, [\href{http://arxiv.org/abs/1012.3751}{{\tt
  1012.3751}}].

\bibitem{laia2011holographic}
J.~N. Laia and D.~Tong, \emph{A holographic flat band}, {\emph{Journal of High
  Energy Physics} {\bf 2011} (2011) 125}.

\bibitem{Alexandrov:2012xe}
V.~Alexandrov and P.~Coleman, \emph{{Spin and holographic metals}},
  \href{http://dx.doi.org/10.1103/PhysRevB.86.125145}{\emph{Phys. Rev. B} {\bf
  86} (2012) 125145}, [\href{http://arxiv.org/abs/1204.6310}{{\tt 1204.6310}}].

\bibitem{Hartnoll:2008vx}
S.~A. Hartnoll, C.~P. Herzog and G.~T. Horowitz, \emph{{Building a Holographic
  Superconductor}},
  \href{http://dx.doi.org/10.1103/PhysRevLett.101.031601}{\emph{Phys.Rev.Lett.}
  {\bf 101} (2008) 031601}, [\href{http://arxiv.org/abs/0803.3295}{{\tt
  0803.3295}}].

\bibitem{Gubser:2008px}
S.~S. Gubser, \emph{{Breaking an Abelian gauge symmetry near a black hole
  horizon}},
  \href{http://dx.doi.org/10.1103/PhysRevD.78.065034}{\emph{Phys.Rev.} {\bf
  D78} (2008) 065034}, [\href{http://arxiv.org/abs/0801.2977}{{\tt
  0801.2977}}].

\bibitem{Armitage:2017cjs}
N.~Armitage, E.~Mele and A.~Vishwanath, \emph{{Weyl and Dirac Semimetals in
  Three Dimensional Solids}},
  \href{http://dx.doi.org/10.1103/RevModPhys.90.015001}{\emph{Rev. Mod. Phys.}
  {\bf 90} (2018) 015001}, [\href{http://arxiv.org/abs/1705.01111}{{\tt
  1705.01111}}].

\bibitem{erlich2005qcd}
J.~Erlich, E.~Katz, D.~T. Son and M.~A. Stephanov, \emph{Qcd and a holographic
  model of hadrons}, {\emph{Physical Review Letters} {\bf 95} (2005) 261602}.

\bibitem{oh2019holographic}
E.~Oh and S.-J. Sin, \emph{Holographic abelian higgs model and the linear
  confinement}, {\emph{arXiv preprint arXiv:1909.13801} (2019) }.

\bibitem{bhatt1982scaling}
R.~N. Bhatt and P.~Lee, \emph{Scaling studies of highly disordered spin-$1/2$
  antiferromagnetic systems}, {\emph{Physical Review Letters} {\bf 48} (1982)
  344}.

\bibitem{paalanen1988thermodynamic}
M.~Paalanen, J.~Graebner, R.~N. Bhatt and S.~Sachdev, \emph{Thermodynamic
  behavior near a metal-insulator transition}, {\emph{Physical review letters}
  {\bf 61} (1988) 597}.

\bibitem{PhysRevLett.48.344}
R.~N. Bhatt and P.~A. Lee, \emph{Scaling studies of highly disordered
  spin-\textonehalf{} antiferromagnetic systems},
  \href{http://dx.doi.org/10.1103/PhysRevLett.48.344}{\emph{Phys. Rev. Lett.}
  {\bf 48} (Feb, 1982) 344--347}.

\bibitem{guo1994quantum}
M.~Guo, R.~N. Bhatt and D.~A. Huse, \emph{Quantum critical behavior of a
  three-dimensional ising spin glass in a transverse magnetic field},
  {\emph{Physical review letters} {\bf 72} (1994) 4137}.

\bibitem{PhysRevX.8.031028}
I.~Kimchi, A.~Nahum and T.~Senthil, \emph{Valence bonds in random quantum
  magnets: Theory and application to ${\mathrm{ybmggao}}_{4}$},
  \href{http://dx.doi.org/10.1103/PhysRevX.8.031028}{\emph{Phys. Rev. X} {\bf
  8} (Jul, 2018) 031028}.

\bibitem{PhysRevB.98.054422}
M.~Watanabe, N.~Kurita, H.~Tanaka, W.~Ueno, K.~Matsui and T.~Goto,
  \emph{Valence-bond-glass state with a singlet gap in the spin-$\frac{1}{2}$
  square-lattice random ${J}_{1}\text{\ensuremath{-}}{J}_{2}$ heisenberg
  antiferromagnet
  ${\mathrm{sr}}_{2}{\mathrm{cute}}_{1\ensuremath{-}x}{\mathrm{w}}_{x}{\mathrm{o}}_{6}$},
  \href{http://dx.doi.org/10.1103/PhysRevB.98.054422}{\emph{Phys. Rev. B} {\bf
  98} (Aug, 2018) 054422}.

\bibitem{Uematsu_2018}
K.~Uematsu and H.~Kawamura, \emph{Randomness-induced quantum spin liquid
  behavior in the s=12 random j1−j2 heisenberg antiferromagnet on the square
  lattice}, \href{http://dx.doi.org/10.1103/physrevb.98.134427}{\emph{Physical
  Review B} {\bf 98} (Oct, 2018) }.

\bibitem{Liu_2018}
L.~Liu, H.~Shao, Y.-C. Lin, W.~Guo and A.~W. Sandvik, \emph{Random-singlet
  phase in disordered two-dimensional quantum magnets},
  \href{http://dx.doi.org/10.1103/physrevx.8.041040}{\emph{Physical Review X}
  {\bf 8} (Dec, 2018) }.

\bibitem{PhysRevLett.123.087201}
K.~Uematsu and H.~Kawamura, \emph{Randomness-induced quantum spin liquid
  behavior in the $s=1/2$ random-bond heisenberg antiferromagnet on the
  pyrochlore lattice},
  \href{http://dx.doi.org/10.1103/PhysRevLett.123.087201}{\emph{Phys. Rev.
  Lett.} {\bf 123} (Aug, 2019) 087201}.

\bibitem{Kawamura_2019}
H.~Kawamura and K.~Uematsu, \emph{Nature of the randomness-induced quantum spin
  liquids in two dimensions},
  \href{http://dx.doi.org/10.1088/1361-648x/ab400c}{\emph{Journal of Physics:
  Condensed Matter} {\bf 31} (sep, 2019) 504003}.

\bibitem{Im}
e.~Im, \emph{Pseudo-gaps and the condensation of singlets in a degenerately
  doped silicon metal}, {\emph{submitted to nature} (2020) }.

\bibitem{Seo:2018hrc}
Y.~Seo, G.~Song, Y.-H. Qi and S.-J. Sin, \emph{{Mott transition with
  Holographic Spectral function}},  \href{http://arxiv.org/abs/1803.01864}{{\tt
  1803.01864}}.

\bibitem{Fu_2015}
M.~Fu, T.~Imai, T.-H. Han and Y.~S. Lee, \emph{Evidence for a gapped
  spin-liquid ground state in a kagome heisenberg antiferromagnet},
  \href{http://dx.doi.org/10.1126/science.aab2120}{\emph{Science} {\bf 350}
  (Nov, 2015) 655--658}.

\bibitem{Oh:2018wfn}
E.~Oh and S.-J. Sin, \emph{{Entanglement String and Spin Liquid with
  Holographic Duality}},  \href{http://arxiv.org/abs/1811.07299}{{\tt
  1811.07299}}.

\end{thebibliography}\endgroup

\end{document}